\documentclass[aps,prb,superscriptaddress,twocolumn,citeautoscript]{revtex4-1}
\usepackage{graphicx}
\usepackage{amsfonts}
\usepackage{amsmath}
\usepackage{amssymb}
\usepackage{mathtools}
\usepackage{bm}
\usepackage{textcomp}

\usepackage{bbold}
\usepackage{mathtools}
\usepackage{dcolumn}
\usepackage{hyperref}
\usepackage[usenames]{color}
\usepackage{subfig}

\usepackage{physics}
\usepackage{units}

\usepackage{tablefootnote}

\usepackage{xcolor}

\usepackage{ragged2e}
\DeclareCaptionJustification{justified}{\justifying}
\def\beq{\begin{equation}}
\def\eeq{\end{equation}}
\def\vk{\textit{\textbf{k}}}
\def\vq{\textit{\textbf{q}}}
\def\vR{\textit{\textbf{R}}}
\def\vM{\textit{\textbf{M}}}
\def\J32{$J_{\mathrm{eff}}=3/2$}

\newcommand{\out}[1]{{}}

\newcommand{\corr}[1]{\textcolor{black}{#1}}
\begin{document}

\title{Multipolar interactions and magnetic excitation gap in d$^3$ spin-orbit Mott insulators}

\author{Leonid V. Pourovskii}
\email{leonid@cpht.polytechnique.fr}
\affiliation{CPHT, CNRS, \'Ecole polytechnique, Institut Polytechnique de Paris, 91120 Palaiseau, France}
\affiliation{Coll\`ege de France, Universit\'e PSL, 11 place Marcelin Berthelot, 75005 Paris, France}

\begin{abstract}
 In Mott insulators with a half-filled $t_{2g}$ shell the Hund's rule coupling induces a spin-3/2 orbital-singlet ground state. The spin-orbit interaction is  not expected to qualitatively
 % lead to important qualitative effects.
  impact low-energy degrees of freedom in such systems. Indeed, $d^3$  cubic double perovskites (DP) of heavy transition metals are believed to exhibit conventional collinear magnetic orders. However, their inelastic neutron scattering spectra feature large  gaps of unclear origin. Here we derive first-principles low-energy Hamiltonians for the cubic DP  Ba$_2$Y$B'$O$_6$ ($B'=$ Os, Ru) and show that they include significant multipolar -- dipole-octupolar -- intesite exchange terms. These terms break continuous symmetry of the spin-3/2 Hamiltonian opening an excitation gap. The calculated gap magnitudes are in good agreement with experiment. The dipole-octupolar intersite exchange is 
 %by analytical calculations 
induced due to excited states of the  $t_{2g}^3$ manifold  that are admixed  by the spin-orbit interaction into the spin-3/2 ground state. 
\end{abstract}

\date{\today}

\maketitle

\section{Introduction} 

Mott insulators of heavy transition metals (TM) exhibit a rich variety of unusual inter-site interactions and ordered phases~\cite{Witczak-Krempa2014,Takayama2021}, like Kitaev physics in $d^5$ irridates \cite{Jackeli2009}, multipolar orders~\cite{Balents2010,Chen_Balents2011,Lu2017,Maharaj2020,Hirai2020,Paramekanti2020,Pourovskii_d2,Khaliulin2021} and valence-bond glasses~\cite{deVries2010,Romhanyi2017} in $d^1$ and $d^2$  DP, or excitonic magnets in $d^4$ perovskites ~\cite{Jain2017}. These exciting phenomena originate in large spin-orbit (SO) entangling the orbital momentum $L$ with spin $S$  thus splitting the ground state (GS)  $LS$ multiplet. The  resulting SO GS is then characterized by the total (pseudo-)angular momentum $J_{\mathrm{eff}}$  that depends on the $d$-shell occupancy %and crystal-field symmetry, 
and determines the space of low-energy local degrees of freedom \cite{Takayama2021}.

The physics of $d^3$ Mott insulators is expected to be more conventional and less interesting. In the presence of a large octahedral or tetrahedral ligand field, the $t_{2g}$ shell is half-filled. The Hund's rule thus forces  $J_{\mathrm{eff}}=S=3/2$  and $L=$0, i.~e. a spin-3/2 orbital singlet GS.  The local TM moments are then, to a first approximation, spins-3/2 with their coupling described by a gapless  isotropic Heisenberg model.  
Excited $t_{2g}^3$ states are separated  by a large Hund's rule gap \cite{Sugano_book} and perturbatively  admixed  into the spin-3/2 GS 
by SO. 
No remarkable qualitative effects have been theoretically shown to stem from this admixture.  
In contrast to the exotic orders of the spin-entangled SO Mott insulators, the $d^3$ systems usually exhibit conventional antiferromagnetism (AFM). In particular, for a number of  $d^3$  DP with the formula $A_2BB'O_6$,  where $B'$ is a heavy magnetic TM, a simple  collinear type-I AFM has been inferred from neutron diffraction~\cite{Battle1989,Carlo2013,Kermarrec2015,Taylor2016,Thompson2016}.  

All these $d^3$ DP systems feature, however, surprisingly ubiquitous large gaps in their inelastic neutron scattering (INS) spectra \cite{Carlo2013,Kermarrec2015,Taylor2016,Maharaj2018,Paddison2023}. The gaps are found in monoclinic DP as well as in the cubic DP Ba$_2$YOsO$_6$ (BYOO) and Ba$_2$YRuO$_6$ (BYRO). In the monoclinic case, an excitation gap could be explained by a single-ion anisotropy induced by the spin-orbit admixture  
to the spin-3/2 GS. Its origin is much less clear in the cubic systems, where, for the GS 
%$\Gamma_8$ 
quadruplet, the single-ion anisotropy  is negligible~\cite{Liu2022}, but the measured excitation gap, $\sim$17~meV in BYOO  and 5~meV in BYRO \cite{Carlo2013,Kermarrec2015}, is still large. 
%No consistent theoretical explanation for the emergence of this gap has been proposed so far. 
 The observed gaps can be fitted by tetragonal single-ion or 2-ion anisotropy terms~\cite{Taylor2016,Maharaj2018}, which are, however, not consistent with the absence of any distortions of the cubic symmetry. In all measured  systems, the gap is consistently several times larger in the 5$d$ system as compared to its 4$d$ equivalent. In BYOO, a significant SO admixture into the $d^3$ GS was confirmed with X-ray scattering by Taylor {\it et al.}~\cite{Taylor2017}. They suggested this admixture to  induce the observed excitation gap without  providing a concrete physical mechanism relating them. 

In this work, we calculate low-energy effective Hamiltonians for BYOO and BYRO in the framework of density functional+dynamical mean-field theory (DFT+DMFT)~\cite{Georges1996,Anisimov1997_1,Lichtenstein_LDApp,Aichhorn2016} by using an  ab initio  force-theorem (FT) method \cite{Pourovskii2016}. These calculations predict unexpectedly large multipolar -- dipole-octupolar (DO) -- intersite exchange interactions (IEI)  that lift a continuous symmetry of the Hamiltonian thus opening  an excitation gap. Our calculation also predict, for both compounds, a non-collinear 2\vk\ transverse magnetic order, which is consistent with the propagation vector detected by neutron diffraction. The calculated INS intensities reproduce the experimental spin gap in BYOO as well as its significant reduction in BYRO. These ab initio results are supported by analytical calculations within a simplified tight-binding model %of $d^3$ DP 
predicting 
leading multipolar IEI to be of the DO type and
%the DO IE
 to scale as a square of SO coupling strength. 
Overall the present theory provides a consistent explanation for the excitation gaps in cubic $d^3$ SO Mott insulators; the same mechanism is shown to enhance the gap in lower symmetry phases.

\begin{figure*}[!bth]
	\begin{centering}
		\includegraphics[width=2.1\columnwidth]{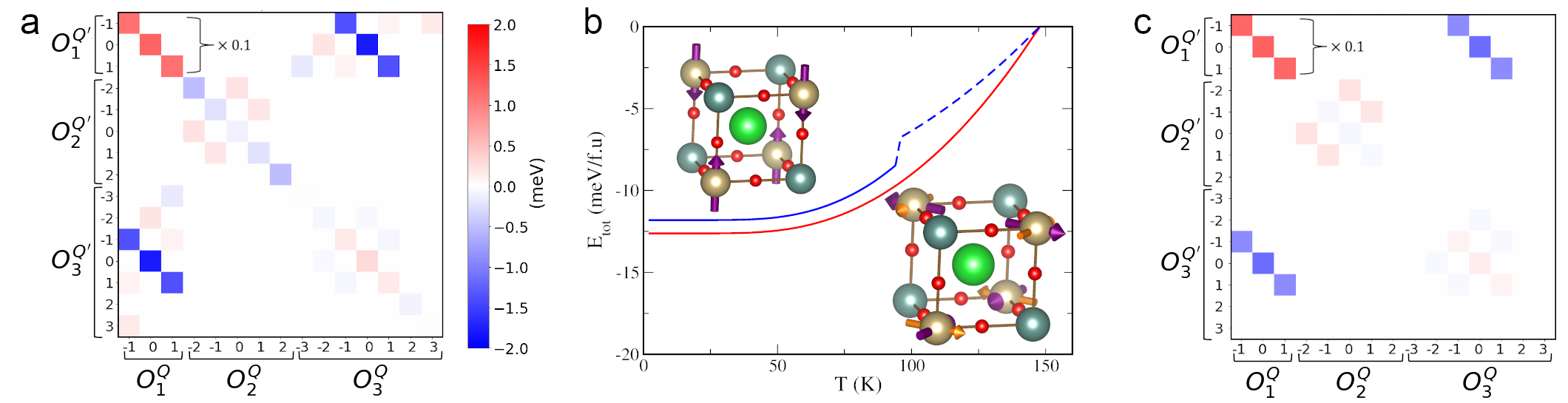}
		\par\end{centering}
	\caption{a. Color map of the ab initio IEI matrix $V_{KK'}^{QQ'}$ for the [1/2,1/2,0] Os-Os pair in BYOO. The dipole-dipole interactions are scaled down by  0.1 in order not to mask other IEI. %One sees that the next largest IEI are ferro dipole-octupolar ones.  
		(b) BYOO mean-field total energy vs. temperature calculated from ab initio $H_{IEI}$ is shown for the
		%planar non-collinear 
		2\vk-P  (lower right corner) and LC (upper left corner) AFM structures by red and blue curves, respectively. In the structure cartoons (plotted by VESTA \cite{Momma2011}), the light brown, turquoise, green and small red balls are Os, Y, Ba and O atoms; the directions of dipole and $\Gamma_5$ octupole moments are shown by the thick purple and thinner orange arrows, respectively. The first-order discontinuity in the blue curve is a spin-flop transition to the TC  AFM ($\vM||[110]$); the part of the curve corresponding to this structure is dashed.  c. Map of the IEI matrix calculated from the analytical superexchange Hamiltonian. The color scale is the same as in panel a.}
	\label{fig1} 
\end{figure*}

The paper is organized as follows. In Sec.~\ref{sec:method} we briefly introduce our ab initio approach (with a detailed description provided in the Appendix). In Sec~\ref{sec:results} we present the ab initio low-energy effective Hamiltonians  and analyze their structure; we subsequently discuss  the ordered phases and excitation spectra of BYOO and BYRO obtained by solving those Hamiltonains. In Sec.~\ref{sec:SE_model} we introduce a simplified tight-binding model of $d^3$ DP and show how the structure of $t_{2g}$ hopping,  in conjunction with the SO coupling, leads to the emergence of the leading multipolar DO IEI.

\section{Ab initio method} \label{sec:method}

We calculate the electronic structure of BYOO and BYRO using the DFT+DMFT approach of Refs.~\onlinecite{Wien2k,Aichhorn2009,Aichhorn2016}  treating Ru 4$d$ and Os 5$d$ states within 
%DMFT by 
the quasi-atomic Hubbard-I (HI) approximation~\cite{hubbard_1}. 
%A rotationally-invariant 5$d$-shell Coulomb vertex is defined by $F^0=U^d=2.6$~eV and Hund's rule $J_H^d=0.39$~eV; they correspond, for the $t_{2g}$ Kanamori Hamiltonian, to $U=$3.05~eV, which is within the accepted range for 5$d$ DP \cite{Erickson2007,Romhanyi2017,FioreMosca2021}, and $J_H$=0.30~eV inferred for BYOO from measurements in Ref.~\cite{Taylor2017}. For BYRO we employ a larger value, $U^d=$3.6~eV,  to account for stronger localization of 4$d$ states. Our qualitative results are insensitive to varying $J_H$ or employing $t_{2g}$ Hamiltonian instead of full $d$-shell one, see Supplementary Material (SM) \cite{supplmat}. 
From the converged DFT+HI electronic structure we calculate all  IEI between the \J32 pseudospins for first several coordination shells using the  FT-HI method of Ref.~\onlinecite{Pourovskii2016}, analogously to its previous applications to  $d^1$ and $d^2$ DP~\cite{FioreMosca2021,Pourovskii_d2}.  Only nearest-neighbor (NN) IEI are found to be important,  the next-NN ones are almost two orders of magnitude smaller. See Appendix~\ref{sec:abinitio_details} for calculational details and Appendix~\ref{sec:el_struct} for the DFT+HI  electronic structure of BYOO and BYRO.

\section{Results}\label{sec:results}

 \subsection{Low-energy  Hamiltonian} 
 
 IEI  between  \J32 GS quadruplets take the following general form
 \beq\label{eq:HSE_general}
H_{IEI}=\sum_{\langle
	ij\rangle}\sum_{KQK'Q'}V_{KK'}^{QQ'}(ij)O_{KQ}^iO_{K'Q'}^j,
 \eeq
 where the on-site multipolar operator $O_{KQ}^i$ is the normalized Hermitian spherical tensor~\cite{Santini2009} for \J32  of the rank $K=$1,2,3 (for dipoles, quadrupoles, octupoles, respectively) and projection $Q$ acting on the site at the position $\vR_i$.     
 	%We employ 
 	These normalized, $\mathrm{Tr}\left[O_{KQ}\cdot O_{K'Q'}\right]=\delta_{KK'}\delta_{QQ'}$, tensors 
 	%to facilitate the comparison of IEI between multipoles of different rank. They
 	are  identical, apart from normalization prefactors, 
 	to the usual definitions of multipoles in terms of non-normalized polynomials of angular momentum operators, e.~g. $O_{10}\equiv O_z=J_z/\sqrt{5}$, $O_{20}\equiv O_{z^2}=\frac{1}{6}(3J_z^2-J(J+1))$, $O_{30}\equiv O_{z^3}=(5J_z^3-3J(J+1)J_z+J_z)/\sqrt{45}$.  
 %in the right-hand-side 
The  IEI $V_{KK'}^{QQ'}(ij)$ couples the multipoles $KQ$ and $K'Q'$ on two magnetic ($B'$) sites connected by the lattice vector $\vR_{ij}=\vR_j-\vR_i$, the  first sum is over all NN bonds $\langle ij\rangle$ in the lattice.  

%We calculate all IEI for BYOO and BYRO for first several correlated shells using the FT-HI method.  
The 
calculated 
BYOO  IEI matrix $\hat{V}(ij)$ for  
%the 
$\vR_{ij}=$[1/2,1/2,0] 
%bond calculated by the FT-HI method 
is depicted in Fig.~\ref{fig1}a. 
%as a color map
The leading IEI are diagonal AFM dipole-dipole (DD) terms $V_{11}^{aa}$, where  $a=-1,0,1\equiv{y,z,x}$ with %the average magnitude of 11.5~meV and
 an axial anisotropy,  $V_{11}^{zz}>V_{11}^{xx(yy)}$.

A striking feature of  BYOO $\hat{V}(ij)$ is unexpectedly large DO terms. 
%The most important feature of the calculated BYOO  IEI matrix is the presence of unexpectedly large dipole-octupolar (DO) terms. 
The leading DO IEI are about 1/8 of the DD ones and ferromagnetic (FM).
% in sign
 Other multipolar IEI are 
 %an order of magnitude 
 at least several times smaller.
% than the leading DO terms
 The picture for BYRO is qualitatively similar to that for BYOO. 
 however, while its DD IEI average 
 of 9.3~meV 
 is close to that in BYOO, both the DD axial anisotropy and DO IEI are an order of magnitude smaller (all calculated IEI for the both systems are listed in Appendix~\ref{sec:IEI_table}).
 % than in the Os system.

 The large DO coupling  takes a simple form for the $xy$ bond,   $\sum_{Q=-1..1}V_{13}^{QQ} O_{1Q}^iO_{3Q}^j$, see Fig.~\ref{fig1}a,  but is less symmetric in the $yz$ and $xz$ planes. %(see SM~\cite{supplmat}).  
 We thus introduce  the operators $\tilde{O}_{KQ}=O_{KQ}/\langle J_{\mathrm{eff}};3/2|O_{K0}|J_{\mathrm{eff}};3/2\rangle$  to get rid of normalization coefficients in subsequent results and transform the octupole operators into  
 %Alternatively, transforming to
 symmetry-adapted octupoles belonging to the $\Gamma_2$,  $\Gamma_4$ and $\Gamma_5$ irreducible representations (IREP) \cite{Shiina1997}. %and 
 Keeping only DD and leading DO IEI, one obtains for  the $xy$ bond:
%
%H'_{xy}=\left[-V_{\Gamma_4}^{\perp}\tilde{O}_{\Gamma_4z}^i\tilde{O}_{1z}^j+V_{\Gamma_4}^{||}\left(\tilde{O}_{\Gamma_4x}^i\tilde{O}_{1x}^j +\tilde{O}_{\Gamma_4y}^i\tilde{O}_{1y}^j\right)+\right. \\
%\left.V_{\Gamma_5}\left(\tilde{O}_{\Gamma_5x}^i\tilde{O}_{1x}^j-\tilde{O}_{\Gamma_5y}^i\tilde{O}_{1y}^j\right)+ (i \leftrightarrow j)\right]+\delta V \tilde{O}_{1z}^i\tilde{O}_{1z}^j,
%\beq\label{eq:HDO_reps}
%\begin{multlined}[t]
\begin{align}\label{eq:HDO_reps}
H'_{xy}&= V\bm{\tilde{O}}_1^i \bm{\tilde{O}}_1^j + \delta V \tilde{O}_{1z}^i\tilde{O}_{1z}^j+\left[ V_{\Gamma_4}^{||}\bm{\tilde{O}}_{\Gamma_4}^i\bm{\tilde{O}}_1^j-V_{\Gamma_4}^{\perp}\tilde{O}_{\Gamma_4z}^i\tilde{O}_{1z}^j\right. \notag \\
 &\left. +V_{\Gamma_5}\left(\tilde{O}_{\Gamma_5x}^i\tilde{O}_{1x}^j-\tilde{O}_{\Gamma_5y}^i\tilde{O}_{1y}^j\right)+ (i \leftrightarrow j)\right],
\end{align}
%\end{multlined}
%\eeq
where the DO term is in the square brackets, the octupole operators are labeled by the  IREP subscript, and $\bm{\tilde{O}}$ are 3D vectors of corresponding operators \footnote{In terms of the operators of Ref.~\onlinecite{Shiina1997} : $\tilde{O}_{1a}=2J_{a}/3$,  $\tilde{O}_{\Gamma_4 a}=4T^{\alpha}_{a}/3$ and $\tilde{O}_{\Gamma_5 a}=4T^{\beta}_{a}/3$, $a=x,y,z$.}. Our calculated
values for $V$, $\delta V$, $V_{\Gamma_4}^{\perp}$, $V_{\Gamma_4}^{||}$, and $V_{\Gamma_5}$ in  BYOO are 5.0, 0.40, 0.39, 0.13, and 0.16   meV, respectively (the formulae for converting the IEI in eq.~\ref{eq:HSE_general} to those in \ref{eq:HDO_reps} are given Appendix~\ref{sec:IEI_table}) .
$H'$  or other bonds are given by cyclic permutation of the indices in (\ref{eq:HDO_reps}).  As we show below, the DO IEI are at the origin of the spin gap in spin-orbit $t_{2g}^3$ cubic DP.

\subsection{Magnetic order}

 Subsequently, we employ the calculated ab initio IEI Hamiltonian (Fig.~\ref{fig1}a) to derive, within a mean-field (MF) approximation~\cite{Rotter2004} , magnetic order as a function of temperature (see Appendix~\ref{sec:MF_details} for relevant methodological details). In the both systems we obtain a planar non-collinear 2\vk\  AFM  order (2\vk-P), depicted in Fig.~\ref{fig1}b, as the  GS. The dipole (magnetic) moments in 2\vk-P are  $M_{x(y)} =(M/\sqrt{2})\exp[i\vk_{y(x)}\vR]$, where the propagation vectors \vk$_x=[1,0,0]$ and \vk$_y=[0,1,0]$ in the units of $2\pi/a$. The \vM direction thus flips by 90$^{\circ}$ between adjacent layers.  The calculated Néel temperatures, $T_N$=146~K in BYOO and 108~K in BYRO, are about twice larger than experimental  69 and 47~K respectively~\cite{Kermarrec2015,Carlo2013}; such systematic overestimation by the present MF-based approach was previously observed for other face-centered cubic (fcc) frustrated magnets \cite{Pourovskii2019,Pourovskii2021,Pourovskii_d2}. Another metastable MF solution -- a longitudinal collinear  type-I AFM structure (LC) with \vk=[0,0,1] -- is found in BYOO at zero temperature to be  about 0.8~meV above the 2\vk-P GS.  (Fig.~\ref{fig1}b).

\begin{figure}[!tb]
	\begin{centering}
		\includegraphics[width=1.0\columnwidth]{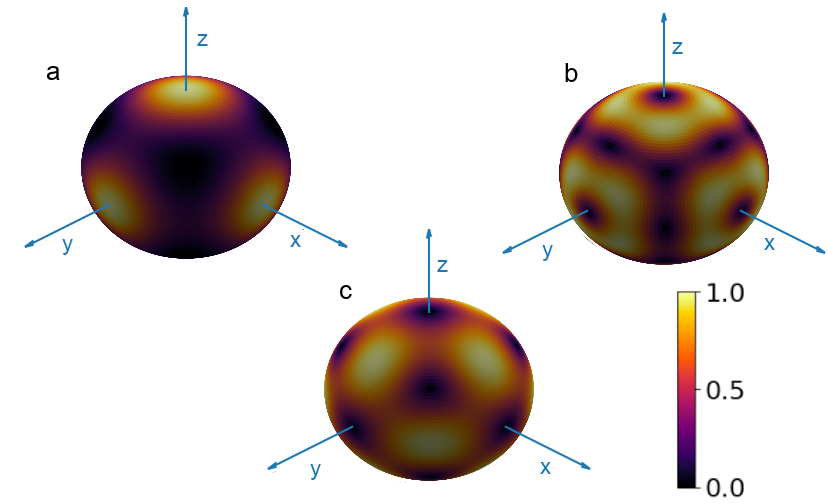}
		\par\end{centering}
	\caption{Magnitude of the octupolar $\Gamma_4$ and $\Gamma_5$ moments as a function of the saturated dipole moment direction $\bm{\hat{M}}=\bm{M}/M$. 
		%The magnitude is indicated by color increasing from dark to light. 
		 a).  $\Gamma_4$ component along $\bm{M}$,  
		 %i.~e. 
		  $|\langle\bm{\tilde{O}}_{\Gamma_4}\rangle\cdot\bm{\hat{M}}|$; b). $\Gamma_4$ component orthogonal to $\vM$, 
		  % i.~e.  
		  $|\langle\bm{\tilde{O}}_{\Gamma_4}\rangle\times\bm{\hat{M}}|;$ c) $|\langle\bm{\tilde{O}}_{\Gamma_5}\rangle|$, $\Gamma_5$ octupole is always orthogonal to $\bm{M}$. 
	 }
	\label{fig2} 
\end{figure}

Experimentally, a  {\it transverse}  collinear type-I AFM   structure (TC) with  \vk=[0,0,1] and moments lying in the $xy$ plane was initially assigned  to both BYRO and BYOO by neutron diffraction \cite{Battle1989,Carlo2013,Kermarrec2015}, though the exact  order type -- single-\vk\ vs. multi-\vk\ -- is still debated \cite{Fang2019,Paddison2023}. The predicted GS 2\vk-P order cannot be distinguished from the TC one on the basis of neutron diffraction only, since the both structures are transverse and feature propagation vectors of the same star. The 1\vk\ LC order is not consistent %\cite{Gaulin_privcomm} 
with  (100) magnetic Bragg peak observed  in the both compounds  \cite{Carlo2013,Kermarrec2015}. 

One may estimate MF total energies for these competing structures, LC, TC and 2\vk-P, which are degenerate in an  isotropic Heisenberg model, by keeping only the leading anisotropic IEI terms (\ref{eq:HDO_reps}). Assuming fully saturated dipole moments  \footnote{The largest eigenvalue eigenstate of $-\bm{\hat{M}}\bm{O}_1$ has  the saturated dipole moment along the unit vector $\bm{\hat{M}}$.} in all structures, we find the anisotropic contribution to MF total energy (per f.u.) of $2\delta V-4V_{\Gamma_4}^{\perp}-4V_{\Gamma_4}^{||}$,  $-2\delta V+4V_{\Gamma_4}^{\perp}-4V_{\Gamma_4}^{||}$, $-2\delta V+4V_{\Gamma_5}$ and $-2\delta V-8V_{\Gamma_5}$ for LC, TC with $\vM||$[100], TC with $\vM||$[110], and 2\vk-P, respectively.

 The DD IEI alone thus leave TC and 2\vk-P degenerated, while LC is penalized by the $\delta V$ term due  an FM alignment of the out-of-plane moments in the $xy$ plane (Fig.~\ref{fig1}b). With the DO terms included, the FM  coupling for $\Gamma_4$  out-of-plane moments favors, in contrast, the LC order. Finally, the  non-collinear 2\vk-P GS is stabilized by $\Gamma_5$ DO coupling. Notice, that the  $\Gamma_5$ moment is always orthogonal to the saturated dipole one and reaches its maximum for the $\left<110\right>$ dipole-moment direction (see Fig.~\ref{fig2}c). Hence, the $\Gamma_5$ DO IEI tend to favor  $90^{\circ}$ angles between dipole moments that are oriented along $\left<110\right>$.  % The MF total energy difference $E(\mathrm{LC})-E($2\vk-P$)=4\delta V+8V_{\Gamma_5} -4V_{\Gamma_4}^{\perp}$ is 1.28~meV/(f.u.)  in BYOO, in good agreement with numerical results obtained with the full Hamiltonian (Fig.~\ref{fig1}b). 
 The GS magnetic structure in $d^3$ cubic DP is thus determined by a delicate balance between the DD IEI anisotropy and DO coupling.

   For the 1\vk\ metastable state we obtain a 1st order transition  LC$\to$TC at $T\approx 0.65 T_N$ (Fig.~\ref{fig2}b). The difference in MF free-energy between   2\vk-P and high-$T$ TC is then rather small (see SM~\cite{supplmat} Sec.~III) and may be affected by beyond-MF corrections. One may thus  suggest that  this TC 1\vk\ order sets in at $T_N$, with a 1st order transition from TC to 2\vk-P at lower $T$. Such a 1st-order transition below $T_N$   was experimentally observed in the order-parameter evolution of BYOO \cite{Kermarrec2015}.
   
   \begin{figure*}[!tbh]
   	\begin{centering}
   		\includegraphics[width=2.1\columnwidth]{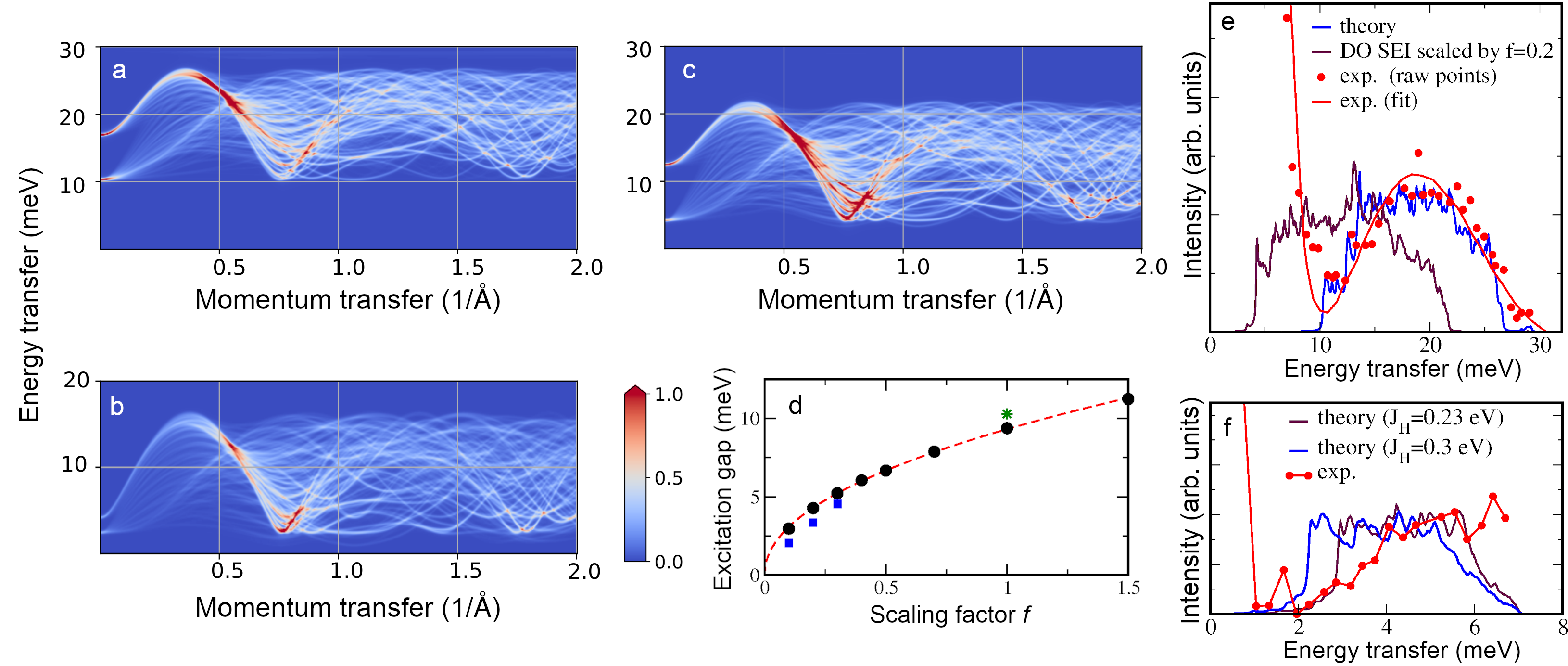}
   		\par\end{centering}
   	\caption{Spherically-averaged INS intensity $S(|\vq|,E)$ calculated from the ab initio IEI for a. BYOO; b. BYRO; c. BYOO $S(|\vq|,E)$ calculated from  the simplified Hamiltonian (\ref{eq:HDO_reps}) with DO IEI scaled down by $f$=0.2; d. Excitation gap at $|\vq|=0.75$~1/\AA\ as a function of scaling factor $f$ for DO IEI in  (\ref{eq:HDO_reps}). Red dashed line is the $\propto f^{0.48}$ fit to the onset of main spectral weight (circles), the blue squares are the position of a  weak resonance appearing at small $f$; the star is the gap value calculated with the full ab initio IEI ; e. Calculated BYOO INS intensity for the initial neutron energy $E_i=$120~meV integrated over  the $|\vq|$ range [0.5:1.5]~1/\AA\ compared to the corresponding experimental data \cite{exp_int_note} from Ref.~\onlinecite{Kermarrec2015}.  f. Calculated BYRO INS intensity for $E_i=$11~meV integrated over  the $|\vq|$ range [0.6:0.9]~1/\AA\ compared to the corresponding experimental data \cite{exp_int_note} from Ref.~\onlinecite{Carlo2013}. }
   	\label{fig3} 
   \end{figure*}

We note that the DO IEI  lift any degeneracy between ordered states that are related by a continuous rotation of dipole moments. The $\Gamma_4$ and $\Gamma_5$ (as well as $\Gamma_2$) octupole moments possess only discrete cubic symmetry. As one sees in Fig.~\ref{fig2}, the dipole moment rotation leads to a change in the relative magnitude of associated $\Gamma_4$ and $\Gamma_5$ octupoles, since they are mixed by any  rotation that is not a cubic symmetry operation. Their IE couplings to dipoles (\ref{eq:HDO_reps}) are distinct and not related by any symmetry in a cubic crystal, therefore, such rotation will  change the energy of dipole order. E.~g., with only anisotropic DD terms included, the TC $\vk=[0,0,1]$ orders are degenerate with respect to a rotation of the ordered moment in the $xy$ plane. With DO IEI   (\ref{eq:HDO_reps})  included, rotating from $\vM||[100]$ to $\vM||[110]$ induces a $\Gamma_5$ octupole and diminishes the $\Gamma_4$ one, thus leading to a change in the DO contribution to the ordering energy. This property of DO IEI has profound implications for magnetic excitations, as shown below.

\subsection{Magnetic excitations} 

We calculate  the INS intensity $S(\vq,E)$ of the 2\vk-P GS using an approach previously applied to $d^2$ DP in Ref.~\onlinecite{Pourovskii_d2}. Namely, the dynamical susceptibility $\chi(\vq, E)$   is calculated in RPA~\cite{RareEarthMag_book} for the MF GS;  the  zero-temperature INS intensity  is then obtained through the fluctuation-dissipation theorem (see, e.~g.,  Ref.~\onlinecite{Lovesey_book_full})  as 
\beq\label{eq:INS_Xsec}
	S(\vq,E)=  \sum_{ab}q^{\perp}_{ab} \\ 
	 \sum_{\mu\mu'\tau\tau'}F_{a\mu}(\vq)F_{b\mu'}(\vq) \mathrm{Im} \chi_{\mu\mu'}^{\tau\tau'}(\vq,E),
\eeq
%\begin{multline}\label{eq:INS_Xsec}
%	\frac{d^2 \sigma}{d \Omega dE'}\propto 
%	S(\vq,E)= \sum_{\alpha\beta}\left(\delta_{\alpha\beta}-q_{\alpha}q_{\beta}\right) \\ \left[\sum_{\mu\mu'\tau\tau'}F_{\alpha\mu}(\vq)F_{\beta\mu'}(\vq) \mathrm{Im} \chi_{\mu\mu'}^{\tau\tau'}(\vq,E)\right],
%\end{multline}
where  $q^{\perp}_{ab}=\delta_{ab}-\hat{q}_{a}\hat{q}_{b}$, $\tau$ and $\mu\equiv KQ$ label sites in the magnetic unit cell and  multipoles, respectively \footnote{In  $S(\vq,E)$ in eq.~\ref{eq:INS_Xsec} we omit   the prefactor $\sqrt{\frac{E_i-E}{E_i}}$ depending on the initial neutron energy $E_i$ in experiment. The prefactor is reinstated in our $|\vq|$-integrated INS spectra in order to have  a quantitative comparison with particular  measurements.}. See Appendix~\ref{sec:INS_details} for further details.

\begin{figure}[!tb]
	\begin{centering}
		\includegraphics[width=0.9\columnwidth]{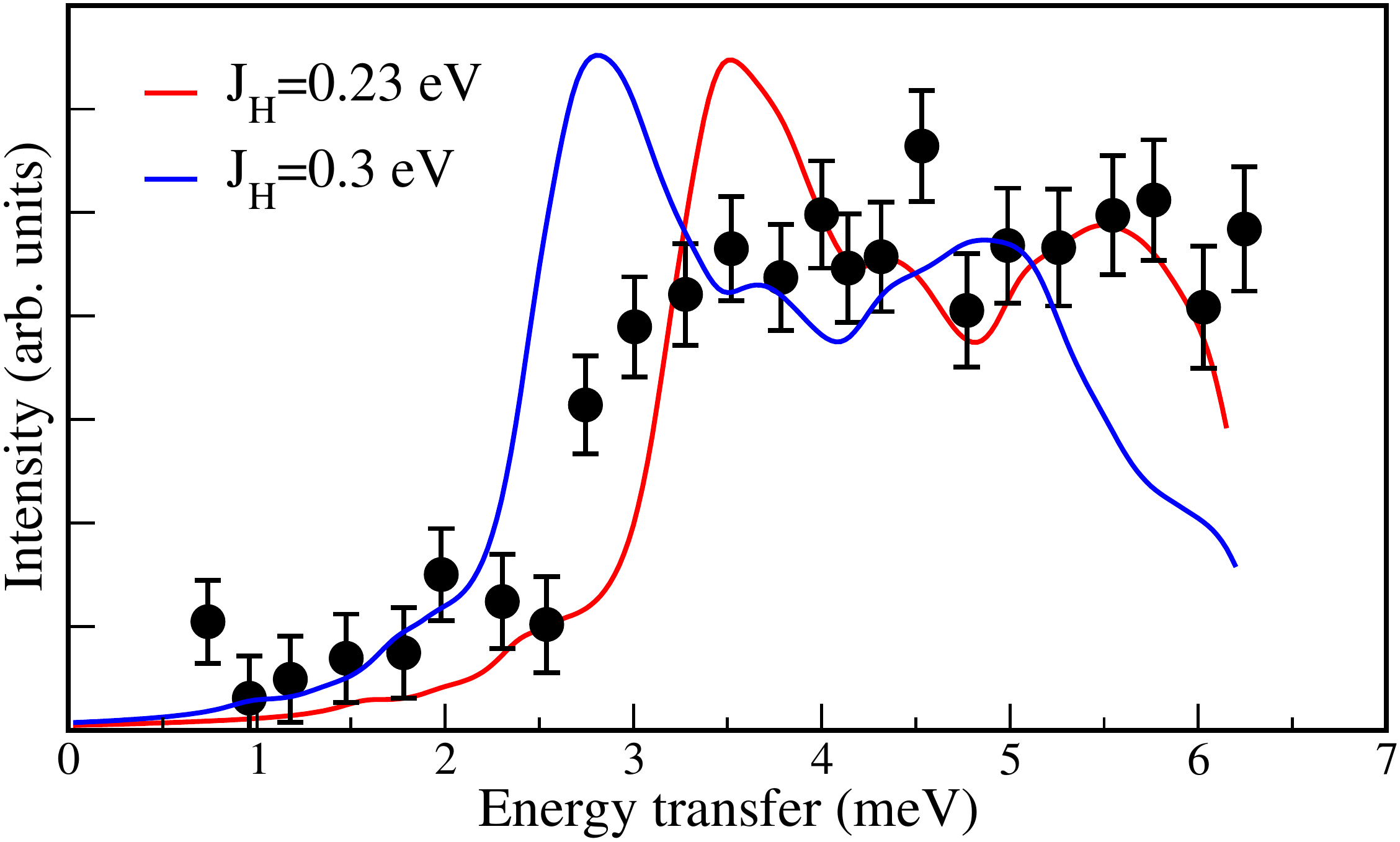} 
		\par\end{centering}
	\caption{Calculated INS intensity in BYRO integrated around  $|\vq|$=0.75~\AA$^{-1}$ (in the range $|\vq|$=[0.7275:0.7625]~\AA$^{-1}$) together with the corresponding experimental data from Ref.~\onlinecite{Paddison2023}. The theoretical curves are calculated for the experimental neutron energy of 11.8 meV and convoluted with a Lorentzian with the width of 0.27~meV corresponding to the experimental instrumental resolution. The experimental error bars are estimated from Fig. 4b of Ref.~\onlinecite{Paddison2023}.   }
	\label{fig:BYRO_INS_vs_JH} 
\end{figure}

The  spherically-averaged INS intensities $S(|\vq|,E)$ for BYOO and BYRO calculated in the  2\vk-P GS structure using the ab initio IEI %. 
%In the both systems one observes a clear excitation gap, with the onset of excitations at about 10~meV and 1.5~meV, respectively. 
exhibit a clear excitation gap (Figs.~\ref{fig3}a and \ref{fig3}b).
In Figs.~\ref{fig3}e and \ref{fig3}f we compare our theoretical $|\vq|$-integrated INS  intensities with  experimental low-temperature ones ~\cite{Kermarrec2015,Carlo2013}  employing the same $|\vq|$-integration  ranges  as in those works \footnote{As reported in Fig.~8 of Ref.~\cite{Kermarrec2015} for BYOO and Fig.~3 of Ref.~\cite{Carlo2013} for BYRO}. 
 We find a nearly perfect quantitative agreement for BYOO. 
 
 The excitation gap is somewhat underestimated in BYRO with  Hund's rule coupling $J_H=0.3$~eV  that we adopted for both compounds; using smaller $J_H$=0.23~eV we  obtain a good agreement for the gap. In  Fig.~\ref{fig:BYRO_INS_vs_JH} we compare the INS intensity for BYRO integrated around $|\vq|$=0.75 \AA$^{-1}$ with very recent experimental data from Ref.~\onlinecite{Paddison2023}. The comparison is displayed for two choices of $J_H$,  0.3~eV    and 0.23~eV. %The value of $U$ is the same in both calculations. 
  One observes a rather good agreement with experiment, which is overall better for the smaller $J_H$ value.
  The excitation gap is seen to be enhanced with decreasing $J_H$ due to the corresponding enhancement of the DO IEI as $\sim (\lambda/J_H)^2$, where $\lambda$ is the SO coupling parameter, see Sec.~\ref{sec:SE_model} below.

 Thus the experimental picture -- of a large excitation gap in these cubic DP with its magnitude being several times larger in BYOO as compared to BYRO -- is fully reproduced by the present theory. 
We note that the position of low-energy intensity peak in the vicinity of the (100) Bragg reflection, $|\vq|=$0.75~\AA$^{-1}$,  is also reproduced in both compounds; the high-energy intensity peak at $|\vq|$ about 0.5~\AA$^{-1}$ is outside of the experimental range in Refs.~\onlinecite{Kermarrec2015,Carlo2013}.   

%The quoted value  values for the excitation gap in BYOO and BYRO, 17 and 5~meV, are in fact the peak intensities of low-energi 

The origin of this excitation gap is DO IEI, which break continuous rotation symmetry of the intersite exchange Hamiltonian, leading to disappearance of Goldstone modes. To demonstrate this explicitly, we employ the simplified Hamiltonian $H'$ (\ref{eq:HDO_reps}) of BYOO with the DO IEI scaled  by a factor $f$. The 2\vk-P GS is stable in the $f$ range  we  explore. At $f$=1 the gap value calculated with $H'$ is very close to that obtained with the full Hamiltonian (\ref{eq:HSE_general}), see Fig.~\ref{fig3}d.   With  $f$=0.2, the gap is reduced to about 4 meV as compared to
%with the onset of excitations shifted
 $\approx$10 meV at $f$=1 (Fig~\ref{fig3}c and Fig~\ref{fig3}e). We carried out this calculation for a set of $f$ values; the resulting gap magnitude   (Fig.~\ref{fig3}d) exhibits a power dependence $\propto f^{\alpha}$, where $\alpha \approx 0.5$. A very weak resonance also appears below the onset of main spectral weight at $f<$0.5, see SM~\cite{supplmat} Sec. VI for details. With the gap scaling as a square-root  of the DO IEI strength $f$  and the latter behaving as  $f\sim (\lambda/J_H)^2$, one finds $\sim 1/J_H$ dependence for the gap; this agrees with the numerical results for BYRO displayed in Fig.~\ref{fig:BYRO_INS_vs_JH}. 

%We have evaluated the impact of DO IEI on the BYOO excitation gap  in the LC magnetic structure (Fig.~\ref{fig1}b) stabilized  by tetragonal distortions (see SM \cite{supplmat} for details). 
%A tetragonal compression $\epsilon_t>$0.5\% is predicted  by our calculations to stabilize it against 2\vk-P due to an easy-axis single-site anisotropy. With yet  larger compression, $\epsilon_t=$1\%, the LC structure is stable even with DO IEI put to zero. 
%Calculating the LC excitation spectra of this tetragonal structure with and without the DO IEI block, we find that the DO IEI double the size of excitation gap. Therefore, DO IEI might be expected to provide a major contribution into the excitation gap in non-cubic spin-orbit d$^3$ Mott insulators as well. 

%===========model analysis of intersite exchange

\section{Dipole-octupolar intersite exchange in a tight-binding model}   \label{sec:SE_model}

%% from SM
In order to clarify the origin of large DO IEI terms, in this section  we derive superexchange interactions in a simplified tight-binding model  relevant for the SO DP. %In the main text these calculations are described rather briefly; here we present the same derivation in more details.

We start with analyzing the impact of SO on the GS of a $t_{2g}^3$ shell.
%we first analyze the impact of SO on the %$J_{\mathrm{eff}}=3/2$
%\J32 states. 
In the absence of SO, the Hund's rule coupling splits 20 states of  the $t_{2g}^3$ manifold into 3 energy levels, which are the GS $^4A_{2}$ quadruplet, 10 degenerate levels belonging to a $^2E$ quadruplet and a $^2T_{1}$ sextet, and an upper $^2T_{2}$ sextet. The  energies of two excited levels are 3$J_{\mathrm{H}}$ and 5$J_{\mathrm{H}}$, respectively, with respect to the GS \cite{Sugano_book}. All these wave functions are listed as Slater determinants in Ref.~\onlinecite{Sugano_book}. Introducing the notation $x,y,z\equiv yz,xz,xy$ for the $t_{2g}$ orbitals, one may write the $^2T_2$ states in the second-quantization notation as
\begin{gather}\label{T2_orbs}
	|^2T_2(x);1/2\rangle =\frac{1}{\sqrt{2}}\left(y_{\downarrow}^{\dagger}z_{\uparrow}-z_{\downarrow}^{\dagger}y_{\uparrow}\right)|{}^4A_{2};3/2\rangle, \nonumber \\
	|^2T_2(y);1/2\rangle=\frac{1}{\sqrt{2}}\left(z_{\downarrow}^{\dagger}x_{\uparrow}-x_{\downarrow}^{\dagger}z_{\uparrow}\right)|{}^4A_{2};3/2\rangle, \\
	|^2T_2(z);1/2\rangle=\frac{1}{\sqrt{2}}\left(x_{\downarrow}^{\dagger}y_{\uparrow}-y_{\downarrow}^{\dagger}x_{\uparrow}\right)|{}^4A_{2};3/2\rangle, \nonumber
\end{gather}
where  $\left|^2T_2(a);1/2\right\rangle$ are the $^2T_{2}$  wavefunctions for the orbital projection $a=x,y,z$ and spin projection $M=\frac{1}{2}$;  $|^4A_{2};3/2\rangle$ is the wavefuntion of the GS quadruplet with $M=\frac{3}{2}$. We also introduced the corresponding creation/annihilation operators for each one-electron orbital $x$, $y$, $z$ and spin.

The SO operator for the $t_{2g}$ shell is $-\lambda\sum_{i}\bm{l}_i\bm{s}_i$, where the SO coupling parameter $\lambda > 0$.   The spin-off-diagonal (spin lowering) part of this operator reads
$$
-\frac{\lambda}{2}l_{+}s_{-}=\frac{\lambda}{2}\left[ (x_{\downarrow}^{\dagger}z_{\uparrow}-z_{\downarrow}^{\dagger}x_{\uparrow})+i(y_{\downarrow}^{\dagger}z_{\uparrow}-z_{\downarrow}^{\dagger}y_{\uparrow})\right],$$
where  $s_{-/+}(l_{-/+})$ is the spin(pseudo-orbital) lowering/raising ladder operator.

Hence, one sees that  in the 1st-order perturbation theory (PT), the SO coupling admixes $^2T_{2}$ states to the pseudo-spin quadruplet, leading to the following expression for the $M=\frac{3}{2}$ state:
\begin{align}\label{eq:pert_GS}
\left|J_{\mathrm{eff}};\frac{3}{2}\right\rangle =&\left|^4A_{2},\frac{3}{2}\right\rangle +  \nonumber\\ & \frac{\epsilon_{\mathrm{SO}}}{\sqrt{2}}\left(\left|^2T_2(y);\frac{1}{2}\right\rangle-i\left|^2T_2(x);\frac{1}{2}\right\rangle\right),
\end{align}
where $\epsilon_{\mathrm{SO}}=\lambda/(5J_\mathrm{H})$. Other $J_{\mathrm{eff}}=3/2$ quartet states are obtained from (\ref{eq:pert_GS}) by a successive application of the $j_{-}=s_{-}-l_{-}$ operator.%, where  $s_{-}(l_{-})$ is the spin(pseudo-orbital)  ladder operator. 

By directly diagonalizing the self-consistent  DFT+HI $t_{2g}$ Os atomic Hamiltonian, we obtain the GS state with  the largest SO admixture from $^2T_2$, but also non-negligible contributions of two other IREP.  Hence, other excited levels, which contribute in the 2nd-order PT ($\sim\epsilon_{\mathrm{SO}}^2$), also admix non-negligibly to the  \J32 GS. The normalized GS quadruplet states read
\beq\label{eq:GS_SO}
\left|J_{\mathrm{eff}};M\right\rangle=\sum_{R \in \mathrm{IREP}}C[R]\left|R;M\right\rangle,
\eeq 
where $C[R]$ is the total contribution due to a given IREP $R$. With the numerical diagonalization  (in which  the ab initio value of $\lambda$=0.294~eV), we obtain the exited level admixtures $C[^2E]$=0.052, $C[^2T_1]$=0.063, and $C[^2T_2]$=0.220, compared to the  1st-order PT result shown above, with only  $C[^2T_2]=\left(1+\epsilon_{\mathrm{SO}}^{-2}\right)^{-1/2}=$0.192 being non-zero. Our magnitudes for the admixture  of excited $t_{2g}^3$ levels to the Os 5d$^3$ GS agree well with estimations  from RIXS measurements \cite{Taylor2017}. As shown below, the 2nd-order $^2E_g$ contribution to the GS is crucial for the DO IEI. 

Subsequently, we employ the GS wavefunctions (\ref{eq:GS_SO}) to calculate BYOO superexchange (SE) analytically within a simplified tight-binding model for the hopping. We  assume the hopping $H_{12}$ between Os $t_{2g}$ shells 1 and 2 that are connected by the $\vR$=[1/2,1/2,0] fcc lattice vector to be given by $\sum_{\sigma}t'\left(x^{\dagger}_{1\sigma}y_{2\sigma}+x^{\dagger}_{2\sigma}y_{1\sigma}\right)-tz^{\dagger}_{1\sigma}z_{2\sigma}+h.c.$, see, e.~g., Ref.~\onlinecite{Takayama2021}. The hopping $t$ between the orbitals ($z$) that lie in the bond plane is dominating, $t>t'$. We further simplify analytical calculations by assuming the same energy for all two-site atomic excitations, $E_0($d$^2$d$^4)=\bar{U}$. Though the latter approximation is rather crude quantitatively, it does not affect qualitative conclusions with respect to the origin of multipolar IEI. The model SE Hamiltonian is then given by $H_{\mathrm{SE}}=-H_{12}^2/\bar{U}=H_{t't'}+H_{tt}+H_{tt'}$, where the three terms  in RHS arise due to the hopping involving only out-of-plane ($x$,$y$) orbitals ($t't'$) , only in-plane ($z$) orbitals ($tt$) and their mixture ($tt'$).  Omitting unimportant single-site contributions, $H_{t't'}$ and $H_{tt}$ read
\begin{align}\label{eq:H1}
	H_{t't'}=&\frac{2(t')^2}{\bar{U}}\sum_{\substack{\sigma\sigma' \\ a=x,y}}\left[(a_{2\sigma}^{\dagger}a_{2\sigma'})(\bar{a}_{1\sigma'}^{\dagger}\bar{a}_{1\sigma})\right. 
	+\nonumber\\
	&\left. (\bar{a}_{2\sigma}^{\dagger}a_{2\sigma'})(\bar{a}_{1\sigma'}^{\dagger}a_{1\sigma})\right],
\end{align}
\beq\label{eq:H3}
H_{tt}=\frac{2t^2}{\bar{U}}\left[\sum_{\sigma\sigma'}(z^{\dagger}_{2\sigma}z_{2\sigma'})(z^{\dagger}_{1\sigma'}z_{1\sigma})\right],
\eeq
where $\bar{x}=y$, $\bar{y}=x$. All the terms in $H_{tt}$ an $H_{t't'}$ are seen to have the same general structure, $X_1X_2$, where both onsite operators $X$ in a given term are of the same type (i.~e., spin and orbital diagonal, either spin or orbital off-diagonal, both spin and orbital off-diagonal). The mixed term $H_{tt'}$ does not contribute to leading multipolar IEI in the \J32 space. 

We then calculate all \J32 SE matrix elements $\langle M_1^1;M_2^2|H_{\mathrm{SE}}|M_3^1;M_4^2\rangle$, where the superscript of $M$ is the site label, and convert  them to the coupling $V_{KK'}^{QQ'}(\vR)$ between on-site moments using  eq.~\ref{eq:Mult_V_conversion}. 

In the zeroth order in $\epsilon_{\mathrm{SO}}$, i.~e.  $\left|J_{\mathrm{eff}};M\right\rangle=\left|^4A_{2},M\right\rangle$, one obtains 
%with $H_{SE}$ 
an isotropic AFM Heisenberg coupling \corr{between spins-3/2}, $5J\sum_{Q=x,y,z}O_{1Q}(1)O_{1Q}(2)\equiv J\vec{S}_1\vec{S}_2$, where $J=(4t^2+8(t')^2)/9\bar{U}$. 

In order to evaluate the relative importance of  SO-admixed  excited states for the SE, we calculate the SE matrix elements with the corresponding wavefunctions $\left|R;M\right\rangle$. The largest non-vanishing SE contributions stemming from the SO admixtures  are  of $O(\epsilon_{\mathrm{SO}}^2)$. They are of the types $\langle{}^4A_{2}^1 ;{}^2E^2|H_{SO}|{}^4A_{2}^1;{}^4A_{2}^2\rangle$ and $\langle{}^4A_{2}^1;{}^2T_2^2|H_{SO}|{}^4A_{2}^1;{}^2T_2^2\rangle$, where we omit the $M$ quantum number for brevity. (Note that matrix elements of the type $\langle{}^4A_{2}^1;{}^2T_2^2|H_{SO}|{}^4A_{2}^1;{}^4A_{2}^2\rangle$, which would contribute in $O(\epsilon_{\mathrm{SO}})$, are all zero, since a non-zero matrix element $\langle{}^4A_{2}|X_1|{}^2T_2\rangle\langle {}^4A_{2}|X_2|{}^4A_{2}\rangle$ requires orbitally off-diagonal $X_1$ and orbitally diagonal $X_2$.) The largest $O(\epsilon_{\mathrm{SO}}^2)$ terms are due to $H_{tt}$; they contribute to DO and anisotropic DD IEI. 

The fact that SE contributions like 
\begin{eqnarray}
\langle{}^4A_{2}^1;{}^2E^2|H_{tt}|{}^4A_{2}^1;{}^4A_{2}^2\rangle\propto \nonumber \\
\sum_{\sigma\sigma'}\langle{}^4A_{2}| z^{\dagger}_{\sigma}z_{\sigma'}|{}^4A_{2}\rangle  \langle{}^2E| z^{\dagger}_{\sigma'}z_{\sigma}|{}^4A_{2} \rangle,
\end{eqnarray}
map within the   \J32 space into a  DO coupling can be shown explicitly by expanding those on-site matrices into multipole moments. Namely, with the magnetic quantum number written explicitly, those $4\times4$ matrices  are   $X^{AA}_{MM'}(\sigma\sigma')=\langle ^4A_{2};M| z^{\dagger}_{\sigma}z_{\sigma'}|^4A_{2};M'\rangle$ and 
$X^{EA}_{MM'}(\sigma'\sigma)=\langle ^2E;M| z^{\dagger}_{\sigma'}z_{\sigma}|^4A_{2};M' \rangle$. By expanding them as $X=\sum_{KQ}\mathrm{Tr} [X\cdot O_{KQ}]O_{KQ}$ one finds that the $X^{AA}$ matrices map only to dipole moments, as expected.  In contrast, the $X^{EA}$ ones map, apart from dipoles, also to octupoles and quadrupoles. The contribution of the latter (which would result in a symmetry-forbidden dipole-quadrupole interaction) is canceled out between Hermitian-conjugated terms in $H_{tt}$, hence, only DD and DO SE terms remain. A similar analysis is applicable for the second  $\sim \epsilon_{\mathrm{SO}}^2$ contribution, $\langle ^4A_{2} ;^2T_2|H_{\mathrm{SO}}|^4A_{2}; ^2T_2\rangle$, since the matrices $X^{TT}_{MM'}(\sigma\sigma')=\langle ^2T_2;M| z^{\dagger}_{\sigma}z_{\sigma'}|^2T_{2};M' \rangle$ also map into dipoles and octupoles.

The corresponding matrix elements of $H_{t't'}$ also contribute in $O(\epsilon_{\mathrm{SO}}^2)$ to both the  DO and anisotropic DD couplings, as well as to quadrupole-quadrupole (QQ) ones; these contributions are smaller by the hopping anisotropy factor $(t'/t)^2$  as compared to the $H_{tt}$ ones. Hence, this analysis confirms that in $t_{2g}^3$ SO double perovskites, the DO couplings are expected to be the largest IEI besides the conventional DD ones.

Employing a reasonable set of parameters ($t=$0.1~eV and  $t'=0.3t$, $\bar{U}=$2~eV)  in  and the ab initio GS wavefunctions (\ref{eq:GS_SO}) in the simplified model described above, we obtain 
%the magnitudes of SE coupling displayed in Fig.~\ref{fig1}c. 
the IEI matrix $\hat{V}$ (Fig.~\ref{fig1}c) that is in a good qualitative agreement with the ab initio one (Fig.~\ref{fig1}a). 
The contribution due to the  $^2E$ admixture is dominant determining an axial anisotropy of DD IEI  with $V_{zz} > V_{xx} =V_{yy}$ (the $^2T_2$ contribution favors a planar anisotropy).  The DO IEI are ferro-coupled pairs of the corresponding moments with $Q=$ -1,0,1(=$y,z,x$); they are an order of magnitude smaller than DD IEI. The QQ and octupole-octupole terms are insignificant.
% % end copy from SM

\section{Summary and Outlook} 

In summary, our ab initio calculations of the low-energy effective Hamiltonians in \corr{the} $d^3$ spin-orbit double perovskites Ba$_2$YOsO$_6$ and Ba$_2$YRuO$_6$ predict significant multipolar  intersite exchange interactions (IEI). Such significant multipolar IEI  are quite unexpected in the case a half-filled $t_{2g}^3$ shell. The leading multipolar  IEI are of a dipole-octupole (DO) type. Namely,  they couple the conventional total angular-moment operators $J_a$ ($a=x,y,z$) acting on a magnetic site (Os or Ru) with octopolar operators, which are time-odd cubic polynomials of $J_a$, acting on its nearest-neighbor magnetic sites. The DO IEI lift continuous symmetry of the effective Hamiltonian  resulting in a gaped excitation spectra. The multipolar IEI are thus at the origin of the large excitation gaps that were previously observed in inelastic neutron scattering spectra (INS) of $d^3$ spin-orbit double perovskites \cite{Carlo2013,Kermarrec2015,Paddison2023}. The theoretical INS spectra calculated  from the effective Hamiltonians are in a good quantitative agreement with  those measurements.   These ab initio results are further supported by analysis in the framework of a simplified analytical model, which predicts the DO terms to be leading IEI,  besides the conventional Heisenberg terms, in $d^3$ cubic double perovskites. Usually,  bi-quadratic (quadrupole-quadrupole) IEI $\sim \left(J_a^iJ_{a'}^j\right)^2$ are assumed to be the most significant multipolar IEI in such $d^3$ systems\cite{Fang2019,Paddison2023}. Our results  contradict this assumption. Moreover, the DO IEI are also predicted to stabilize a non-collinear $2\vk$ transverse structure, which propagation vector $\vk=\langle 1,0,0\rangle$ agrees with experiment\cite{Carlo2013,Kermarrec2015} .

On the basis of our analysis, the leading dipole-octupolar  IEI  are  expected to scale as  $(\lambda/J_H)^2$, where $\lambda$ is the spin-orbit coupling strength and $J_H$ is the Hund's rule coupling. Since $J_H$ is weakly changing along the 4$d$ and 5$d$ TM series and between them, the dipole-octupolar IEI magnitude $f$ is effectively controlled by $\lambda^2$. The numerical RPA calculations for the excitation gap  vs. $f$ (Fig.~\ref{fig3}d)  find that the gap scales as $\sqrt{f}\propto \lambda$ thus explaining  the fact that the measured gap in   $5d$ systems is several times larger compared  to that in equi-electronic $4d$ systems. 

Moreover, the DO IEI can also be expected to provide a major contribution to the excitation gap in non-cubic spin-orbit d$^3$ Mott insulators. 
To estimate this contribution, we have also evaluated for  Ba$_2$YOsO$_6$  the excitation gap  in the LC magnetic structure (shown in Fig.~\ref{fig1}b) stabilized  by  1\% of tetragonal compression (see SM \cite{supplmat} Sec. IV for details). 
A tetragonal compression $\epsilon_t>$0.5\% is predicted  by our calculations to stabilize it against 2\vk-P due to an easy-axis single-site anisotropy. With yet  larger compression, $\epsilon_t=$1\%, the LC structure is stable even with DO IEI put to zero. 
Calculating the LC excitation spectra of this tetragonal structure with and without the DO IEI block, we find that the DO IEI double the magnitude of the excitation gap. This  confirms that the effect of DO IEI on the gap is still significant even in systems with a large single-ion anisotropy.

\section*{Acknowledgements}
The author is grateful to B. Gaulin, A. Georges, and  C. Franchini for useful discussions and to the CPHT computer  team for support.

\appendix

\section{Methodological details}\label{sec:app_method}

\subsection{Ab initio calculations}\label{sec:abinitio_details}
Our DFT+HI calculations are based on the full-potential LAPW code Wien2k\cite{Wien2k} and include the SO interaction with the standard second-variation approach. Projective Wannier orbitals \cite{Amadon2008,Aichhorn2009} representing  Os (Ru) $d$ orbitals are constructed from the Kohn-Sham (KS) bands in the energy range [-1.4:4.8] ([-1.4:4.1])~eV relative to the KS Fermi level; this energy window includes all $t_{2g}$ and most of $e_g$ states but not the oxygen 2$p$ bands (see SM\cite{supplmat} for plots of the KS density of states in BYOO and BYRO).

A rotationally invariant Coulomb vertex for the full $d$ shell is  constructed using the parameters $U^d=F^0$ and $J^d_H=(F^2+F^4)/14$ together with the standard additional approximation\cite{Anisimov1993} for the ratio of Slater parameters $F^4/F^2=$0.625.  Some test calculations for BYOO were carried out using "small window" including only Os $t_{2g}$ states and the Kanamori rotationally invariant $t_{2g}$ Hamiltonian with the corresponding parameters  $U=U^d+8J_H^d/5$ and $J_H=0.77 J_H^d$.  In all calculations of BYOO, unless specified otherwise, we employ $F^0=U^d=2.6$~eV and Hund's rule $J_H^d=0.39$~eV. For the $t_{2g}$ Kanamori Hamiltonian they correspond to $U=$3.05~eV, which is within the accepted range for 5$d$ DP \cite{Erickson2007,Romhanyi2017,FioreMosca2021}, and $J_H$=0.30~eV inferred for BYOO from measurements in Ref.~\onlinecite{Taylor2017}. For BYRO, unless noted otherwise, we employ the same value of $J_H$ as in BYOO and the larger value of $U^d=$3.6~eV  to account for a stronger localization of 4$d$ states.

%Some test calculations for BYOO were carried out using "small window" including only Os t$_{2g}$ states and the Kanamori rotationally invariant $t_{2g}$ Hamiltonian with the corresponding parameters  $U=U^d+8J_H^d/5$ and $J_H=0.77 J_H^d$; they lead to very similar results as compared to the full-$d$ shell calculations. 
%The only significant difference concerns the $g$ factor  $J=3/2$ pseudospin, which value  is smaller ($g$=1.64) when calculated using  full 5$d$-shell framework. The reason is that additional  admixture of $e_g$ states into the pseudospin further reduces the weight of main spin-3/2 $^4A_2$ component.    

All calculations are carried out for the experimental cubic lattice structures of BYRO~\cite{Aharen2009} and BYOO \cite{Kermarrec2015}. We employ the local density approximation as the DFT exchange-correlation potential, 400 $\vk$-points in the full Brillouin zone, and the Wien2k basis cutoff $R_{\mathrm{mt}}K_{\mathrm{max}}=$8. The double-counting correction is evaluated using the fully-localized limit with the nominal $d$ shell occupancy of 3.  Extensive benchmarks demonstrate the  robustness of our qualitative results with respect to varying $J_H$ , (see Fig.~\ref{fig:BYRO_INS_vs_JH}), DFT calculational parameters, double-counting correction or employing the $t_{2g}$ Hamiltonian instead of the full $d$-shell, see SM\cite{supplmat} Sec.~II.

%\subsection{Calculations of inter-site exchange interactions (IEI)}
Calculations of IEI $V_{KK'}^{QQ'}(ij)$ acting within the \J32  space are carried out using the FT-HI approach of Ref.~\onlinecite{Pourovskii2016}, analogously to previous applications of  this method to actinide dioxides~\cite{Pourovskii2019,Pourovskii2021} as well as  to d$^1$  and $^2$ double perovskites~\cite{FioreMosca2021,Pourovskii_d2}.  This approach is similar to other magnetic force theorem methods for symmetry-broken phases (Refs.~\onlinecite{Liechtenstein1987,Katsnelson2000}, 
see also
 Ref.~\onlinecite{Szilva2022} for a recent review) 
but is formulated for the paramagnetic state. Within the FT-HI, the matrix elements of IEI   $V(ij)$ coupling \J32 quadruplets on two $B'$ sites  read
\beq\label{V}
\langle M_1 M_3| V(ij)| M_2 M_4\rangle=\mathrm{Tr} \left[ G_{\langle ij\rangle}\frac{\delta\Sigma^{at}_{j}}{\delta \rho^{M_3M_4}_{j}} G_{\langle ji\rangle}\frac{\delta\Sigma^{at}_{i}}{\delta \rho^{M_1M_2}_{i}}\right],
\eeq
where  $<ij>\equiv \vR_j-\vR_i$ is the lattice vector connecting the two sites,  $M=-3/2,...,3/2$ is the magnetic quantum number,  $\rho^{M_1M_2}_{i}$ is the corresponding element of the $J_{\mathrm{eff}}$-quadruplet density matrix on site $i$, $\frac{\delta\Sigma^{at}_{i}}{\delta \rho^{M_1M_2}_{i}}$ is the derivative of atomic (Hubbard-I) self-energy $\Sigma^{at}_{i}$ over a fluctuation of the $\rho^{M_1M_1}_{i}$ element, $G_{\langle ij\rangle}$ is the inter-site Green's function.  The self-energy derivatives are calculated from atomic Green's functions  using analytical formulas derived in Ref.~\onlinecite{Pourovskii2016}, where  the FT-HI method is described in detail. The method is applied  as a post-processing on top of DFT+HI, hence, all quantities in the RHS of eq.~\ref{V} are evaluated from a fully converged DFT+HI electronic structure.

Once all matrix elements (\ref{V}) are calculated, we make use of the orthonormality property $\mathrm{Tr}\left[O_{KQ}\cdot O_{K'Q'}\right]=\delta_{KK'}\delta_{QQ'}$ of the Hermitian multipolar operators $O_{KQ}$ (which are defined in accordance with eq.~10 of Ref.~\onlinecite{Santini2009}) to map them into the IEI $V_{KK'}^{QQ'}(ij)$ between on-site moments:
\begin{gather}\label{eq:Mult_V_conversion}
V^{QQ'}_{KK'}(ij)= \\
\sum_{\substack{M_1M_2 \\ M_3M_4}} \langle M_1M_3| V(ij)|  M_2M_4\rangle \left[O_{KQ}\right]_{M_2M_1} \left[O_{K'Q'}\right]_{M_4M_3}\nonumber.
\end{gather}

To have a correct mapping into the $J_{\mathrm{eff}}$ pseudo-spin basis, the phases of the  $|J_{\mathrm{eff}};M\rangle$ states are chosen such that $\langle J_{\mathrm{eff}};M|J_{+}|J_{\mathrm{eff}};M-1\rangle$ is a positive real number.

\subsection{Mean-field (MF) solution of the effective Hamiltonian}\label{sec:MF_details}

We employ the MCPHASE package~\cite{Rotter2004}  in conjunction with an in-house module implementing  multipolar operators in the MCPHASE framework  to solve the  effective Hamiltonian $H_{IEI} $ in mean-field. %  obtaining stable ordered phases vs. temperature. 
As initial guesses of the MF procedure we employ all  1\vk\  structures realizable within single fcc unit cell; these calculations converge to the 2\vk-P order. In order to obtain a metastable 1\vk\ solution we start with the corresponding initial guess switching off the random Monte Carlo flips implemented in the MCPHASE.  With this procedure the LC structure is obtained at low $T$ independently of whether it or the  TC one is used as the initial guess.

\subsection{Inelastic neutron scattering (INS) intensities} \label{sec:INS_details}

We evaluated the generalized dynamical susceptibility $\chi(\vq,E)$ for the MF  ground state using a generalized random phase approximation (RPA), see Ref.~\onlinecite{RareEarthMag_book}. The INS intensity is  calculated from $\chi(\vq,E)$  by eq.~\ref{eq:INS_Xsec}  using the form-factors $F_{a\mu} (\vq)$  for \J32 multipole  $\mu\equiv {KQ}$, where  $a=x,y,z$. Our approach for evaluating these form-factors  is based on analytical expressions for the one-electron neutron scattering operator $Q_{a}(\vq)$ from Ref.~\onlinecite{Lovesey_book_full}, which matrix elements in the $d^3$ $J_{\mathrm{eff}}$ space  are calculated with the HI eigenstates  of the \J32 quadruplet. The resulting matrices  are then expanded in multipole operators\cite{Shiina2007} as 
$$
\langle J_{\mathrm{eff}};M|Q_{a}(\vq)|J_{\mathrm{eff}};M'\rangle=\sum_{\mu}F_{a\mu}(\vq)\left[O_{\mu}\right]_{MM'}
$$  
to obtain the form factors.

The method  is  described in detail in Supp. Material of Ref.~\onlinecite{Pourovskii_d2}. The radial integrals $\langle j_L(q)\rangle$ for the Os$^{5+}$ 5$d$ shell, which enter into the formulas for one-electron matrix elements of $Q_{a}(\vq)$,  were taken from Ref.~\onlinecite{Kobayashi2011}. For Ru$^{5+}$, the full set of  $\langle j_L(q)\rangle$ has not been given in the literature, to our awareness. We thus use an estimate for Ru$^{5+}$ $\langle j_0(q)\rangle$ from Ref.~\onlinecite{Parkinson2003}; for $L=2,4$  we assume the same values of $\langle j_L(q)\rangle$   as in  Os$^{5+}$.

The spherically averaged INS intensities $S(|\vq|,E)$ are calculated for each $|\vq|$ by averaging over 642 $\vq$-points on an equidistributed icosahedral mesh. 

\section{Electronic structure of BYOO and BYRO}\label{sec:el_struct}

The IEI calculations by the FT-HI method were carried out starting from the converged DFT+HI electronic structure of BYOO and BYRO. 

\begin{figure}[!tb]
	\begin{centering}
		\includegraphics[width=0.9\columnwidth]{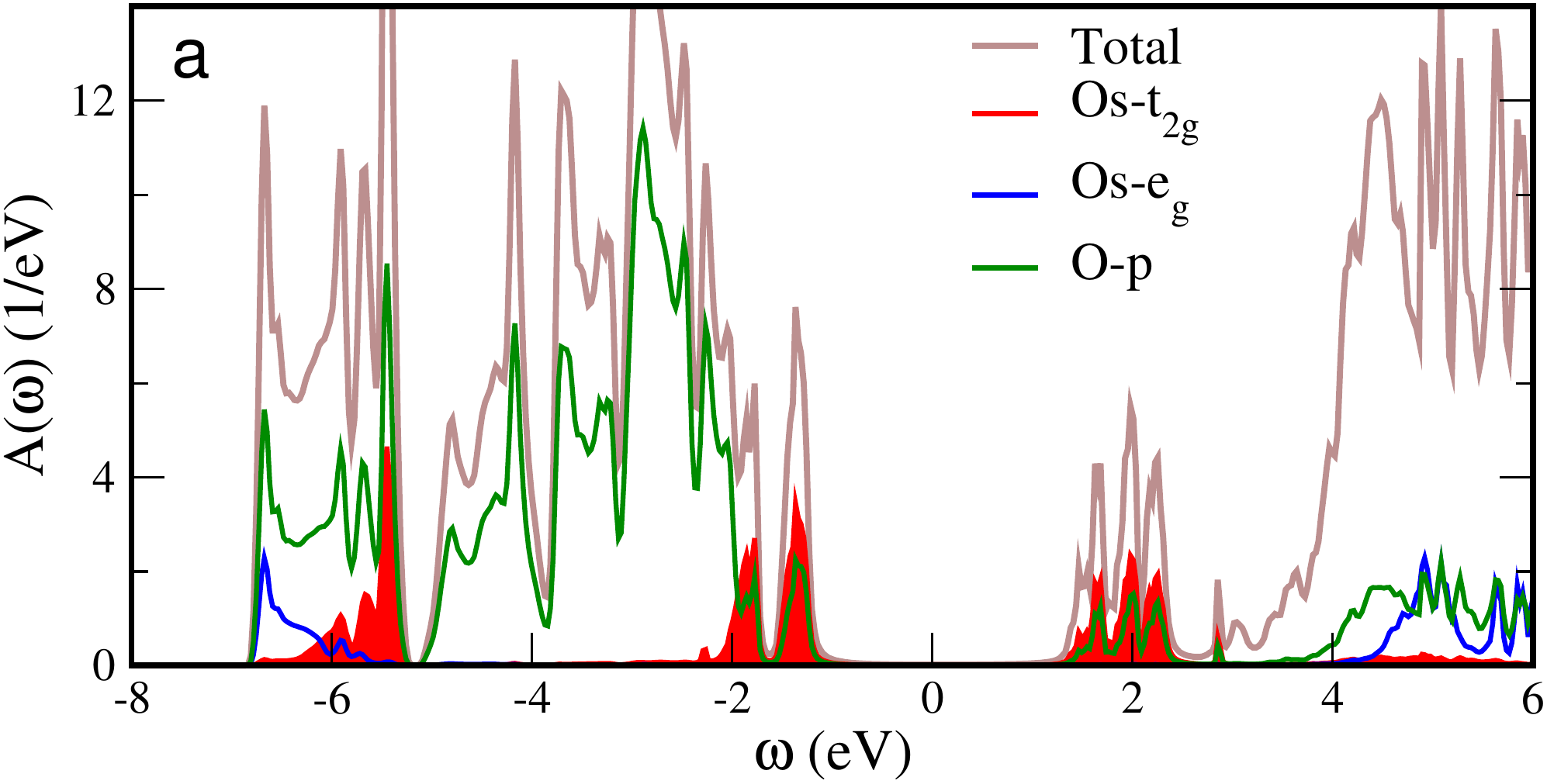}
		\includegraphics[width=0.9\columnwidth]{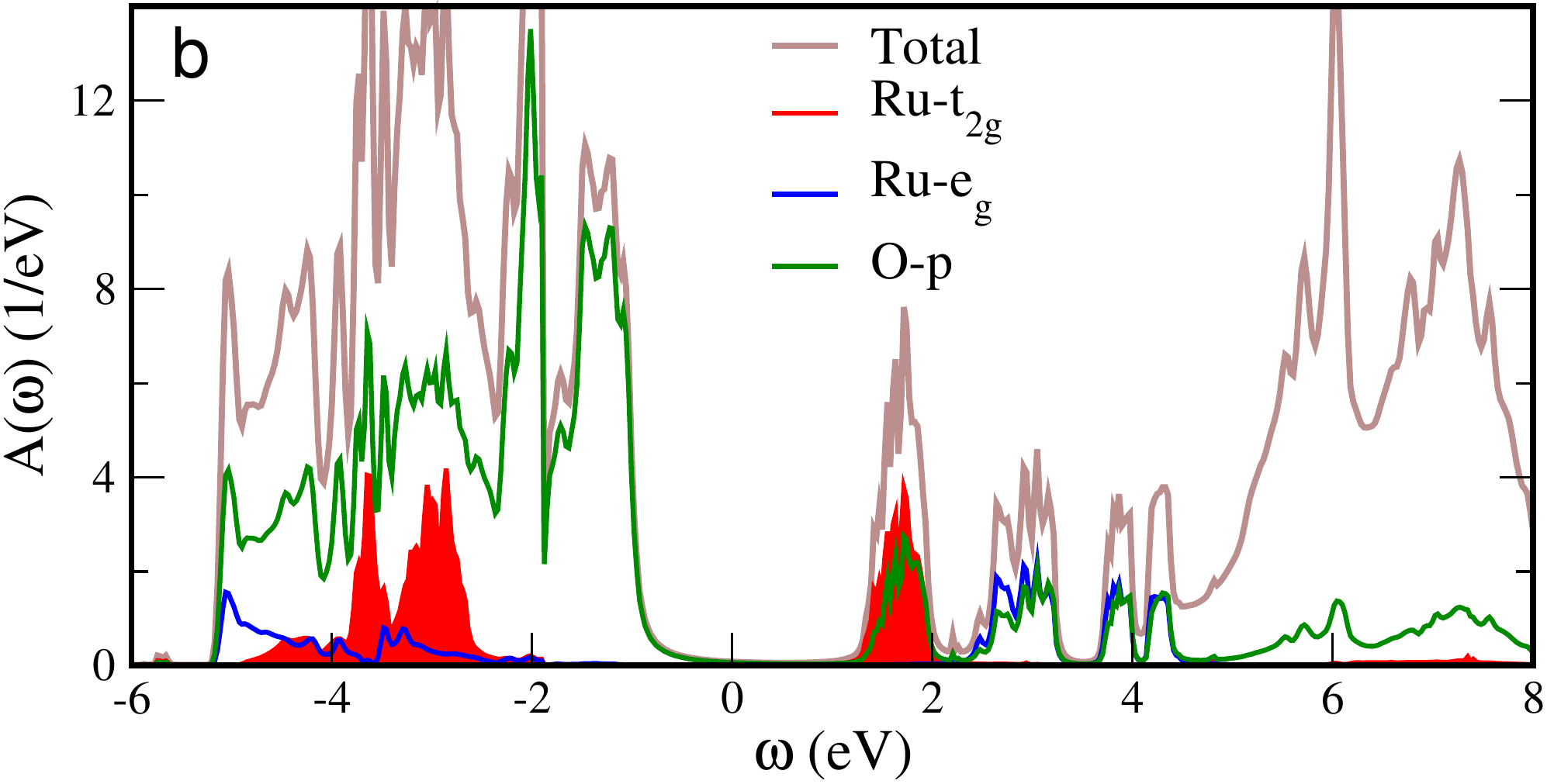}
		\par\end{centering}
	\caption{DFT+HI spectral function of BYOO (a) and BYRO (b). The partial $t_{2g}$ spectral function is shaded in red.}
	\label{fig:HI_DOS} 
\end{figure}

In Fig.~\ref{fig:HI_DOS} we display the converged  DFT+HI spectral functions of the both compounds obtained with the full-$d$ correlated subspace \footnote{In the converged DFT+HI electronic structure the chemical potential is sometimes found to be pinned at the very top of the valence (lower Hubbard) band instead of being strictly inside the Mott gap. This is a drawback of the HI approximation. In those cases, the chemical potential is manually shifted inside the gap.}. Both systems are predicted by DFT+HI to be correlated insulators with the gap of about 2.4~eV and 1.9~eV in BYOO and BYRO respectively. The insulating gap in BYOO is between the Os $t_{2g}$ lower and upper Hubbard bands (HB), hence, this compound is predicted to be a Mott insulator. In contrast, BYRO is a charge-transfer insulator, since the gap is between the upper edge of O 2$p$ valence band and the Ru upper HB. DFT+HI predictions for the gap magnitude are not expected to be quantitatively accurate, since the HB width is known to be underestimated in this approximation~\cite{Kotliar_RMP} leading to the gap being overestimated as noted, e.~g., in the case of rare-earth sesquioxides \cite{Boust2022}.  There are no published experimental data on the gap magnitude or transport in BYOO and BYRO, to our awareness. The DFT+HI electronic structure compares qualitatively well with previous DFT-based calculations (which used somewhat different parameters). In particular, Refs.~\onlinecite{Wang2019,Fang2019} also predicted a Mott gap in BYOO to open between $t_{2g}$ HB, though those calculations had to be made in a magnetically ordered phase due to the well-known limitation of standard DFT(+U) methods in capturing local-moment paramagnetism. They employed smaller values of U, correspondingly,  their calculated  gap was also smaller than the one we find.  Ref.~\onlinecite{Chen2018} employing DFT+U and DFT+DMFT predicted both paramagnetic and antiferromagnetic  BYRO to be insulating
for the values of $U$ and $J_H$ employed in the present work.

For the sake of reproducibility, we also plot the auxiliary non-interacting Kohn-Sham densities of states of BYOO and BYRO in SM\cite{supplmat} Sec.~I.

\section{Intersite exchange interactions}\label{sec:IEI_table}

In Table~\ref{Tab:SEI} we list all calculated IEI  in  BYOO and BYRO  with magnitude above 0.05 meV. The IEI are given for the [0.5,0.5,0.0] nearest-neighbor fcc lattice vector.  

We also list below the formulas to convert these IEI into the IEI of the simplified Hamiltonain (eq.~\ref{eq:HDO_reps}):
\begin{gather}
	V=\frac{9}{20}V_{11}^{11}, \\
	\delta V=\frac{9}{20}(V_{11}^{00}-V_{11}^{11}),\\
	V_{\Gamma_4}^{\perp}=-\frac{3}{20}\left(V_{13}^{00}+\sqrt{\frac{3}{8}}V_{13}^{11}\right), \\
	V_{\Gamma_4}^{||}=-\frac{3}{20}\sqrt{\frac{3}{8}}V_{13}^{11}, \\
	V_{\Gamma_5}=-\frac{3}{20}\sqrt{\frac{5}{8}}V_{13}^{11},
\end{gather}
where the overall prefactors (9/20 and 3/20) are due to the change of operators normalization from $O$ to $\tilde{O}$, and the square-root factors are due to the transformation to the cubic IREP.

\renewcommand\floatpagefraction{0.1}
\begin{table}[tp]
	\caption{\label{Tab:SEI}  
		%\justifying
		Calculated IEI $V_{KK'}^{QQ'}$. First two columns list
		$Q$ and $Q'$ , respectively. Third and fourth columncolumns list Cartesian labels for the $KQ$ and $K'Q'$ tensors. The last three columns display the values of all  IEI for BYOO and BYRO (meV) with  magnitude above 0.05~meV (for BYRO we list IEI  calculated using two values of $J_H$).
	}
	\begin{center}
		\begin{ruledtabular}
			\renewcommand{\arraystretch}{1.2}
			\begin{tabular}{c c c c c c c }
				& & &  & BYOO & BYRO  & BYRO \\
				& &  &  & $J_H$= 0.3 eV & 0.3 eV  & 0.23 eV \\
				\hline
				\hline
				\multicolumn{7}{c}{Dipole-Dipole} \\
				\hline
				-1 & -1   & y  & y  &   11.22  &    9.27  &    9.67  \\
				0 &  0   & z  & z  &   12.12  &    9.34  &    9.79  \\
				1 &  1   & x  & x  &   11.22  &    9.27  &    9.67  \\
				\hline
				\hline
				\multicolumn{7}{c}{Dipole-Octupole} \\
				\hline
				-1 & -1   & y  & yz$^2$  &   -1.38  &   -0.11  &   -0.17  \\
				-1 &  1   & y  & xz$^2$  &    0.10  &     &     \\
				-1 &  3   & y  & x(3x$^2$-y$^2$)  &    0.16  &     &     \\
				0 & -2   & z  & xyz  &    0.21  &     &     \\
				0 &  0   & z  & z$^3$  &   -1.78  &   -0.13  &   -0.21  \\
				1 & -3   & x  & y(x$^2$-3y$^2$)  &   -0.16  &     &     \\
				1 & -1   & x  & yz$^2$  &    0.10  &     &     \\
				1 &  1   & x  & xz$^2$  &   -1.38  &   -0.11  &   -0.17  \\
				\hline
				\hline
				\multicolumn{7}{c}{Quadrupole-Quadrupole} \\
				\hline
				-2 & -2   & xy  & xy  &   -0.50  &     &     \\
				-1 & -1   & yz  & yz  &   -0.22  &     &     \\
				0 & -2   & z$^2$  & xy  &    0.23  &     &     \\
				0 &  0   & z$^2$  & z$^2$  &   -0.10  &     &     \\
				1 & -1   & xz  & yz  &    0.19  &     &     \\
				1 &  1   & xz  & xz  &   -0.22  &     &     \\
				2 &  2   & x$^2$-y$^2$  & x$^2$-y$^2$  &   -0.51  &     &     \\
				\hline
				\hline
				\multicolumn{7}{c}{Octupole-Octupole} \\
				\hline
				-2 & -2   & xyz  & xyz  &   -0.07  &     &     \\
				-1 & -1   & yz$^2$  & yz$^2$  &    0.16  &     &     \\
				0 & -2   & z$^3$  & xyz  &   -0.06  &     &     \\
				0 &  0   & z$^3$  & z$^3$  &    0.28  &     &     \\
				1 & -1   & xz$^2$  & yz$^2$  &   -0.06  &     &     \\
				1 &  1   & xz$^2$  & xz$^2$  &    0.16  &     &     \\
				2 &  2   & z(x$^2$-y$^2$)  & z(x$^2$-y$^2$)  &   -0.09  &     &     \\
			\end{tabular}
		\end{ruledtabular}
	\end{center}
\end{table}

%\bibliography{bibliography}

\begin{thebibliography}{63}
\bibitem [{exp()}]{exp_int_note}%
	\BibitemOpen
	\href@noop {} {}\bibinfo {note} {The experimental curve is vertically shifted
		to set its minimum to 0.}\BibitemShut {Stop}%
\end{thebibliography}


\end{document}

\title{Supplemental Material for 
'Multipolar interactions and magnetic excitation gap in d$^3$ spin-orbit Mott insulators'}

\author{Leonid V. Pourovskii}
%\email{leonid@cpht.polytechnique.fr}
\affiliation{CPHT, CNRS, \'Ecole polytechnique, Institut Polytechnique de Paris, 91120 Palaiseau, France}
\affiliation{Coll\`ege de France, Universit\'e PSL, 11 place Marcelin Berthelot, 75005 Paris, France}

%\date{\today}	
\maketitle

\section{Kohn-Sham electronic structure of BYOO and BYRO}

\begin{figure}[!b]
	\begin{centering}
		\includegraphics[width=0.57\columnwidth]{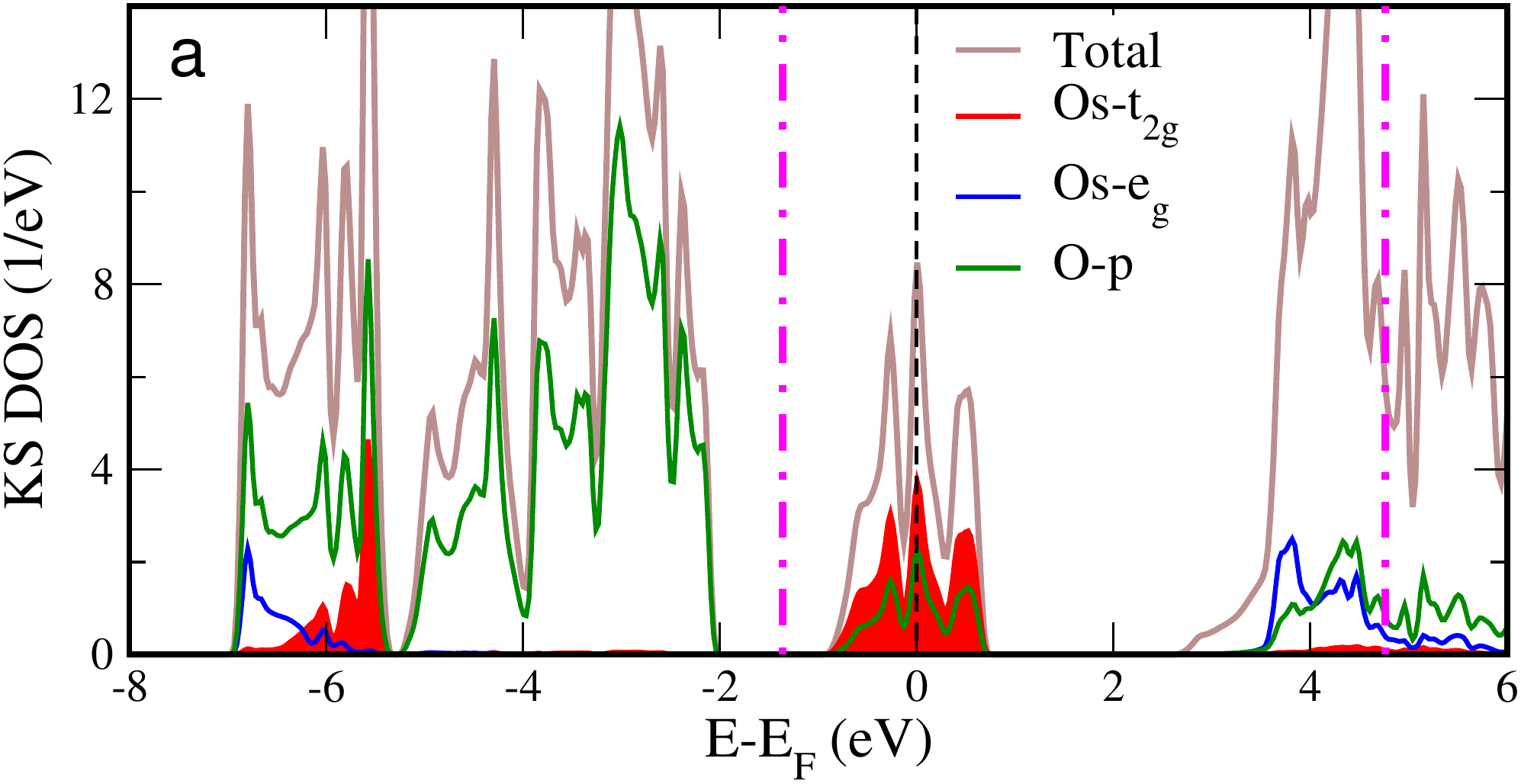}
		\includegraphics[width=0.57\columnwidth]{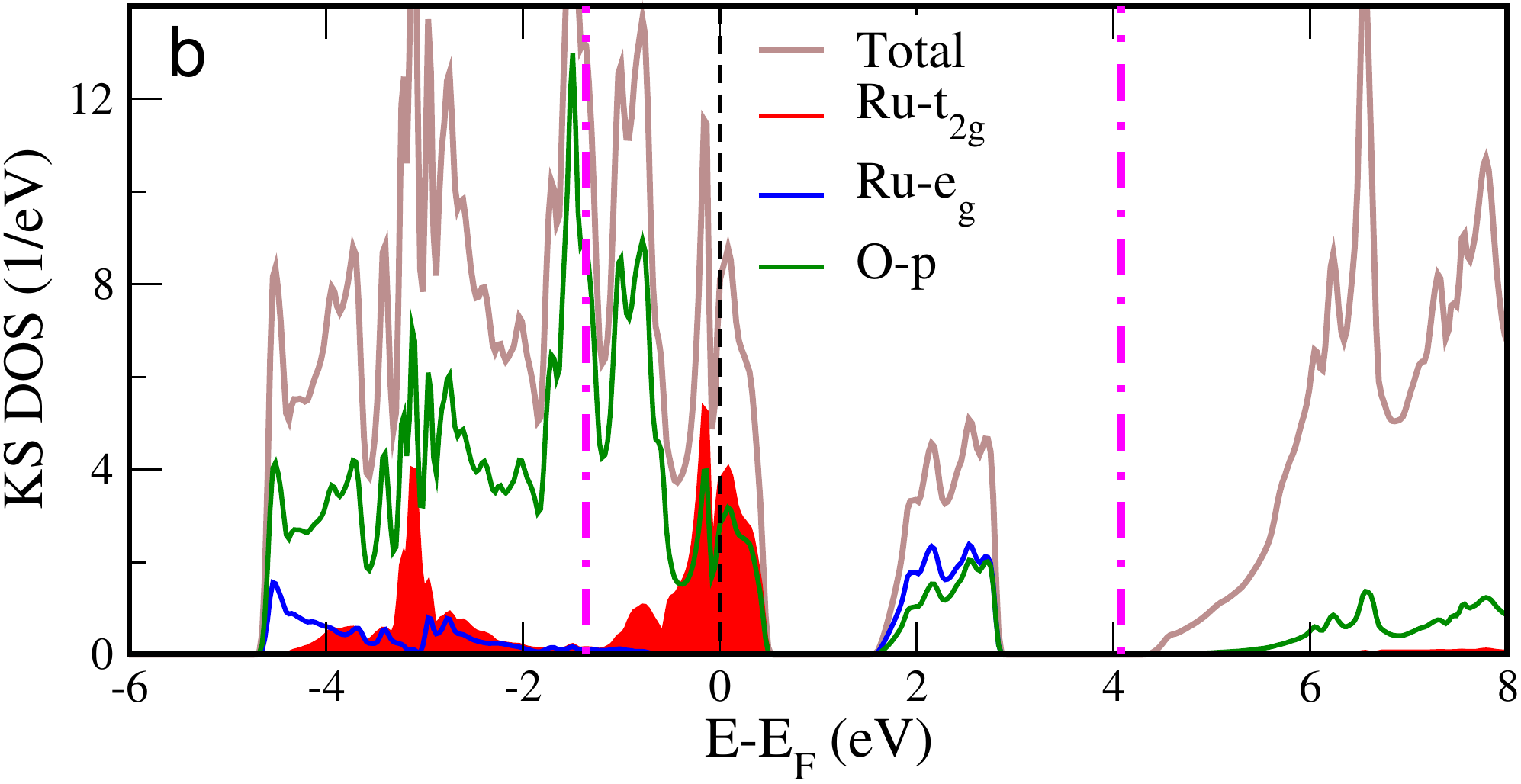}
		\par\end{centering}
	\caption{Kohn-Sham (KS) DOS of BYOO (a) and BYRO (b) as obtained with charge self-consistent DFT+HI calculations. The partial $t_{2g}$ DOS is shaded in red. The magenta vertical dot-dashed lines indicate the window for constructing projective Wannier orbitals for the  full-$d$-shell.}
	\label{fig:KS_DOS} 
\end{figure}

The auxiliary non-interacting Kohn-Sham densities of states (DOS) 	as obtained in  charge self-consistent DFT+HI calculations are displayed in Suppl. Fig.~\ref{fig:KS_DOS}. One sees that the $t_{2g}$ band of the magnetic $B'$(=Os, Ru) ion is pinned at the Fermi level. Overall, the Ru 4$d$ states are located lower in energy as compared to the Os 5$d$ ones. The Ru $t_{2g}$ band overlaps with the top of O 2$p$ band, while the Os $t_{2g}$ one is separated from O 2$p$ by a gap. Correspondingly, the anti-bonding Os $e_g$ band overlaps with other conduction states%(mainly, empty Y 4$d$)
, while the Ru $e_g$ band is separated from the onset of other conduction states by a gap.  The energy window for full-$d$ Wannier construction is chosen to include the $t_{2g}$ band and the $e_g$ antibonding band as indicated in Suppl. Fig.~\ref{fig:KS_DOS}, whereas the O 2$p$ states must be excluded to preserve the correct $t_{2g}$-$e_g$ crystal-field splitting (which arises due to a strong  hybridization of O 2$p$ with $e_g$ states) in the HI effective atomic problem. The $t_{2g}$-only correlated basis, which is used for tests in BYOO, was  naturally constructed including only the separate manifold of Os  $t_{2g}$ bands.

\section{Precision and robustness of calculated IEI}\label{sec:IEI_precision}

The magnitude of calculated IEI (Fig. 1a and Table~I of the main text) is  rather small,  with the crucial dipole-octupole (DO) IEI of about 1 meV only. As shown in the main text, the ground state magnetic structure is determined by a subtle balance between the DO and anisotropic dipole-dipole terms. It is thus important to understand whether our conclusions on the ground state order and on the relative importance of different multipolar IEI are affected by the precision of DFT+HI calculations and the choice of calculational setup.   One may easily demonstrate (as detailed below) that with our choices of calculational parameters  (e.~g.,  the number of \vk-points used in the Brillouin zone integration or the size of LAPW basis set),  the value of DFT+HI  total energy is converged with respect to those parameters to a precision of tens meV only . One may thus wonder whether the IEI calculated by the FT-HI method are well converged with respect to those parameters. Another question is whether the IEI  are qualitatively affected by the DFT+HI calculational setup (the choice of the correlated subspace, of the double-counting correction etc.).

In fact, one may argue that numerical imprecision  in electronic structure calculations affects the IEI in a very different way as compared to the total energy. One may illustrate this difference using the simplest expression for superexchange $J$ in oxides that reads
$$
J \sim \frac{t_{pd}^4}{{(\Delta}_{\mathrm{CT}}^2U)}=\frac{t_{dd}^2}{U},
$$ 
where $t_{pd}$ is the O 2$p$ to TM $d$ hopping, $\Delta_{\mathrm{CT}}$ is the $pd$ charge transfer energy, $t_{dd}$ is the effective hopping between $d$-shell through oxygen (e.~g., Ref.~\onlinecite{Khomskii_book}). The numerical error bars in $J$ thus stem from the errors $\delta t_{pd}$ and $\delta\Delta_{\mathrm{CT}}$  in the $pd$ hopping  and charge transfer energy, respectively. One may easily show that 
$$
\delta J\sim J\left(O(\delta t_{pd} /t_{pd})+O(\delta\Delta_{\mathrm{CT}} /\Delta_{\mathrm{CT}})\right),
$$
i. e. the error in $J$ is given by the relative errors in $t_{pd}$ and $\Delta_{\mathrm{CT}}$ scaled by $J$ itself. For example, if the error in those parameters is 1\%, it will be at most several percent in $J$, however small $J$ is. If $U\gg t_{dd}$ , which is the case in correlated insulators, $J$ is small, and the error will be scaled accordingly.

\renewcommand\floatpagefraction{0.1}
\begin{table}[tp]
	\caption{\label{Tab:benchmarks}  
		%\justifying
		Calculated IEI $V_{KK'}^{QQ'}$ (meV) in BYOO using various calculational parameters and setups. $U^d=$2.6~eV and $J_H=$0.3eV are used throughout. The calculational setup is summarized below for each case. The case 1 is the actual setup employed in the paper. 
	}
	\begin{center}
		\begin{ruledtabular}
			\renewcommand{\arraystretch}{1.4}
			\begin{tabular}{c c c c  c c c c c }
				\multicolumn{4}{c}{{\bf CASE}} &{\bf 1}  & {\bf 2} &  {\bf 3} &  {\bf 4} & {\bf 5} \\
				\multicolumn{4}{c}{Correlated subspace} &	Full-$d$ &	Full-$d$ &	Full-$d$	& Full-$d$ & $t_{2g}$ \\
				\multicolumn{4}{c}{Double-counting formula} & FLL  & FLL &  FLL &  AMF & FLL \\
				\multicolumn{4}{c}{LAPW basis cutoff, $R_{\mathrm{MT}}K_{\mathrm{MAX}}$} & 8  & 7.5 &  7 &   8 & 8 \\
				\multicolumn{4}{c}{Number of \vk-points in the full BZ}  & 400  & 200 &  100 &   400 & 400 \\
				\multicolumn{4}{c}{DFT+HI total energy with respect to {\bf CASE 1} (meV)}& 0  &  61.2 &   328 &     &    \\
				\hline
				\hline
				\multicolumn{9}{c}{Dipole-Octupole} \\
				\hline
				-1 & -1   & y  & y  &   11.22  &   11.23  &   11.45  &   10.53  &    9.53  \\
				0 &  0   & z  & z  &   12.12  &   12.13  &   12.34  &   11.35  &   10.28  \\
				1 &  1   & x  & x  &   11.22  &   11.23  &   11.45  &   10.53  &    9.53  \\
				\hline
				\hline
				\multicolumn{9}{c}{Dipole-Octupole} \\
				\hline
				-1 & -1   & y  & yz$^2$  &   -1.38  &   -1.36  &   -1.40  &   -1.24  &   -1.13  \\
				-1 &  1   & y  & xz$^2$  &    0.10  &    0.10  &    0.11  &    0.10  &    0.09  \\
				-1 &  3   & y  & x(3x$^2$-y$^2$)  &    0.16  &    0.16  &    0.17  &    0.17  &    0.14  \\
				0 & -2   & z  & xyz  &    0.21  &    0.20  &    0.22  &    0.21  &    0.18  \\
				0 &  0   & z  & z$^3$  &   -1.78  &   -1.77  &   -1.81  &   -1.61  &   -1.47  \\
				1 & -3   & x  & y(x$^2$-3y$^2$)  &   -0.16  &   -0.16  &   -0.17  &   -0.17  &   -0.14  \\
				1 & -1   & x  & yz$^2$  &    0.10  &    0.10  &    0.11  &    0.10  &    0.09  \\
				1 &  1   & x  & xz$^2$  &   -1.38  &   -1.36  &   -1.40  &   -1.24  &   -1.13  \\
				\hline
				\hline
				\multicolumn{9}{c}{Quadrupole-Quadrupole} \\
				\hline
				-2 & -2   & xy  & xy  &   -0.50  &   -0.49  &   -0.49  &   -0.51  &   -0.41  \\
				-1 & -1   & yz  & yz  &   -0.22  &   -0.22  &   -0.22  &   -0.23  &   -0.18  \\
				0 & -2   & z$^2$  & xy  &    0.23  &    0.23  &    0.24  &    0.23  &    0.19  \\
				0 &  0   & z$^2$  & z$^2$  &   -0.10  &   -0.10  &   -0.10  &   -0.10  &   -0.08  \\
				1 & -1   & xz  & yz  &    0.19  &    0.18  &    0.19  &    0.19  &    0.15  \\
				1 &  1   & xz  & xz  &   -0.22  &   -0.22  &   -0.22  &   -0.23  &   -0.18  \\
				2 &  2   & x$^2$-y$^2$  & x$^2$-y$^2$  &   -0.51  &   -0.50  &   -0.50  &   -0.52  &   -0.42  \\
				\hline
				\hline
				\multicolumn{9}{c}{Octupole-Octupole} \\
				\hline
				-2 & -2   & xyz  & xyz  &   -0.07  &   -0.07  &   -0.07  &   -0.07  &   -0.06  \\
				-1 & -1   & yz$^2$  & yz$^2$  &    0.16  &    0.15  &    0.16  &    0.14  &    0.12  \\
				0 & -2   & z$^3$  & xyz  &   -0.06  &   -0.06  &   -0.06  &   -0.05  &     \\
				0 &  0   & z$^3$  & z$^3$  &    0.28  &    0.28  &    0.29  &    0.26  &    0.23  \\
				1 & -1   & xz$^2$  & yz$^2$  &   -0.06  &   -0.06  &   -0.07  &   -0.06  &   -0.05  \\
				1 &  1   & xz$^2$  & xz$^2$  &    0.16  &    0.15  &    0.16  &    0.14  &    0.12  \\
				2 &  2   & z(x$^2$-y$^2$)  & z(x$^2$-y$^2$)  &   -0.09  &   -0.08  &   -0.09  &   -0.09  &   -0.07  \\
			\end{tabular}
		\end{ruledtabular}
	\end{center}
\end{table}

In contrast, the total energy will include the $pd$ bonding energy, which in the simplest approximation reads $E_b\sim t_{pd}^2/\Delta_{\mathrm{CT}}$, hence,  $\delta E_b \sim E_b(O(\delta t_{pd} /t_{pd})+O(\delta\Delta_{\mathrm{CT}} /\Delta_{\mathrm{CT}}))$. This expression is similar to the one  for $\delta J$ above, apart from the prefactor $E_b\gg J$. Hence, the same precision in electronic structure calculations will lead to much smaller error bars in the IEI as compared to the total energy.  

To support this qualitative argumentation, we carried out some benchmarks for BYOO using different sets of calculational parameters that determine the precision of DFT+HI total energy. The results of these benchmarks are summarized in Suppl. Table~\ref{Tab:benchmarks}.

Namely, the IEI and total energy were simultaneously calculated varying the LAPW basis-set cutoff\cite{Wien2k} ($R_{\mathrm{mt}}K_{\mathrm{max}}$) and the density of \vk-point mesh. The results are listed in Suppl. Table~\ref{Tab:benchmarks} as the cases 1 to 3. One sees that the reduction of calculational precision with respect to that used in the paper (case 1) has a large impact on the total energy, which exhibits variations that are much larger than the IEI magnitude (about 300 meV/f.u. for the least precise setup). In contrast, the IEI values exhibit only insignificant changes upon the reduction of the DFT+HI precision. Even the smallest IEI are well converged.

One may consider the double-counting (DC) correction as another possible source of uncertainty. The systems under consideration are Mott insulators, for which the standard DC choice is the fully-localized limit as employed in the paper. However, a test calculation using a popular alternative, the around mean-field limit (AMF), has  also been carried out for BYOO and is included in Suppl. Table~\ref{Tab:benchmarks} as case 4. The effect of changing DC on the electronic structure consists in a downwards shift of the Os 5$d$-band  by about 0.4 eV with respect to the O-2p band. The  effect on the IEI consists in a rather uniform reduction of the dipole-dipole (DD) and DO terms by about 10\% without any significant changes in their structure. In contrast, the quadrupole-quadrupole (QQ) IEI remain virtually unchanged between the two cases.  

The reason for this is likely related the structure of superexchange in this system. As shown by analytical calculations for the simplified model (see Sec.~IV of the main text), the same superexchange processes involving the same multiplets of the $t_{2g}^3$ shell give leading contributions to both the DD and DO terms, resulting in a uniform change of the DD and DO blocks upon modifications of the computational setup. In contrast, the leading contributions to the QQ block are different. 

Finally, a test calculation employing the $t_{2g}$  correlated state basis instead of the full-$d$ one was also carried out (case 5). This leads to a more significant (about 20\%), but still uniform reduction of all IEI. The qualitative results are not affected, the same 2\vk-P structure is found in mean-field as the ground state.

\begin{figure}[!tb]
	\begin{centering}
		\includegraphics[width=0.55\columnwidth]{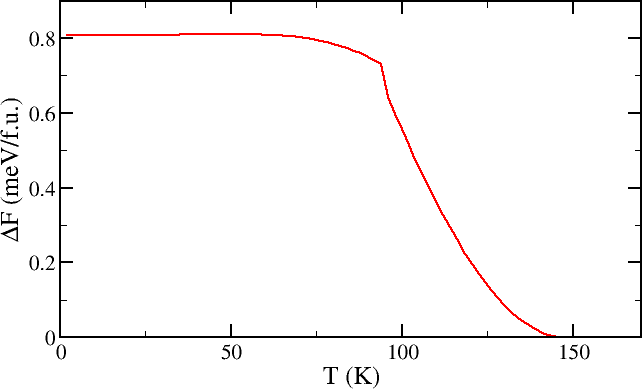}
		\par\end{centering}
	\caption{Difference in the mean-field free energy  between the 2\vk-P ground state and the  metastable 1\vk\  solution.
	}
	\label{fig:Delta_F} 
\end{figure}

\section{Mean-field free energy}

In Suppl. Fig.~\ref{fig:Delta_F} we display the free energy difference $\Delta F=F(1\vk)-F($2\vk-P$)$ between the 2\vk-P GS and the  metastable 1\vk\ solution as a function of temperature, as calculated in MF.  One may notice a change of slope at the first-order transition point of the 1\vk\ structure (Fig.~1b of the main text) resulting in a  reduced energy difference between  2\vk-P and  the high-temperature TC structure.

\section{Magnetic excitations in tetragonally compressed BYOO}

We include the effect of tetragonal compression by adding to the ab initio IEI Hamiltonian (eq. 1 of the main text) a single-ion anisotropy (SIA) term. Hence, we neglect the impact of those distortions on  the IEI, which should be a reasonable approximation at relatively small distortion levels. The SIA term is evaluated within the DFT+HI approach by running calculations for a series of tetragonally compressed BYOO, with the relative distortion $\epsilon_t=(c/a-1)$ ranging from -10$^{-3}$ to -10$^{-2}$. The DFT+HI calculational parameters are the same as in the cubic case and described in Appendix of the main text. 

In result, one obtains splitting of  the \J32 quadruplet by tetragonal crystal field as a function of $\epsilon_t$.  The resulting single-site anisotropy term reads:
\beq\label{eq:HSIA}
H_{\mathrm{SIA}}=\sum_i D \left(J_z^i\right)^2=\sum_i 5D \left(O_{1z}^i\right)^2,
\eeq
where the sum is over all Os sites, $D$ is the SIA coefficient. In the RHS of (\ref{eq:HSIA}) we rewrite the same term using the normalized dipole tensors ($J_z=\sqrt{5}O_{1z}$). The anisotropy coefficient is linear vs. distortion within the studied  range, $D=K\epsilon_t$, where $K=256$~meV. Hence, one sees that tetragonal compression leads to an easy-axis anisotropy and tends to stabilize the longitudinal collinear LC phase against the $2\vk$-P one. By solving $H_{\mathrm{IEI}}+H_{\mathrm{SIA}}$ in MF, we find that the LC phase becomes the magnetic ground state  at the compression level $\epsilon_t\approx - 5\cdot10^{-3}$.  

\begin{figure}[!tb]
	\begin{centering}
		\includegraphics[width=0.62\columnwidth]{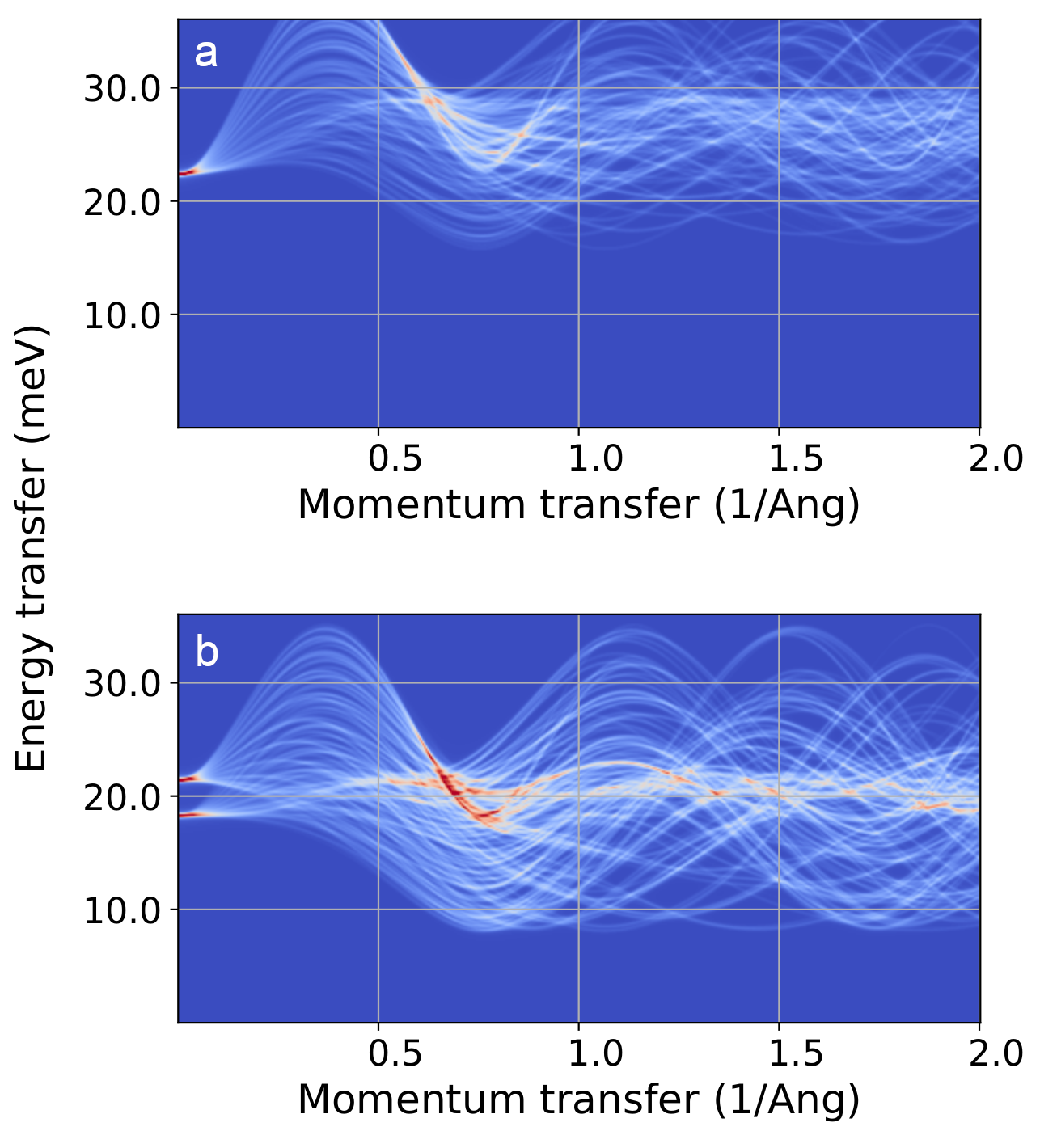} 
		\par\end{centering}
	\caption{Calculated spherically-averaged INS intensity $S(|\vq|,E)$  of tetragonally compressed BYOO at  $\epsilon_t=$-0.01, (a)  with all IEI taken into account and (b) with the DO IEI  put to zero. }
	\label{fig:INS_tetragonal} 
\end{figure}

We then carry out calculations of the magnetic excitation spectra with and without the DO block included.  The corresponding INS intensities  $S(|\vq|,E)$ for $\epsilon_t=-10^{-2}$   (at this compression the LC order is the magnetic GS even without the DO block included) are displayed in Suppl. Fig.~\ref{fig:INS_tetragonal}. The excitation gap is about 8~meV without the DO IEI, i.~e. when it stems from the SIA term only; it is twice larger with the DO term included.

\section{Magnetic excitations in transverse collinear magnetic structure}

  Our {\it ab initio }calculations predict non-collinear 2\vk-P order to be the GS magnetic structure in both BYRO and BYOO. The transverse collinear (TC) order was previously suggested for both BYOO and BYRO  on the basis of neutron diffraction \cite{Carlo2013,Kermarrec2015}. As noted in the main text, it is  not possible to distinguish single-\vk\ and multi-\vk\ orders for the same \vk-vector on the basis of neutron diffrection only; a recent experimental study\cite{Paddison2023} suggests, for example, a 3\vk\ structure for BYRO on the basis  of measured INS compared to excitation spectra calculated from their {\it ad hoc}  magnetic Hamiltonian. It is thus instructive to evaluate the magnetic excitation spectra for the TC structure as  well, for the sake of comparison.
  
  In order to stabilize the TC magnetic order against the LC and 2\vk-P ones in the cubic structure, we employed the simplified Hamiltonian (eq. 2 of the main text), in which the sign of either $V_{\Gamma_5}$ or  $V_{\Gamma_4}^{\perp}$ IEI was flipped.
  
    \begin{figure}[!tb]
  	\begin{centering}
  		\includegraphics[width=0.57\columnwidth]{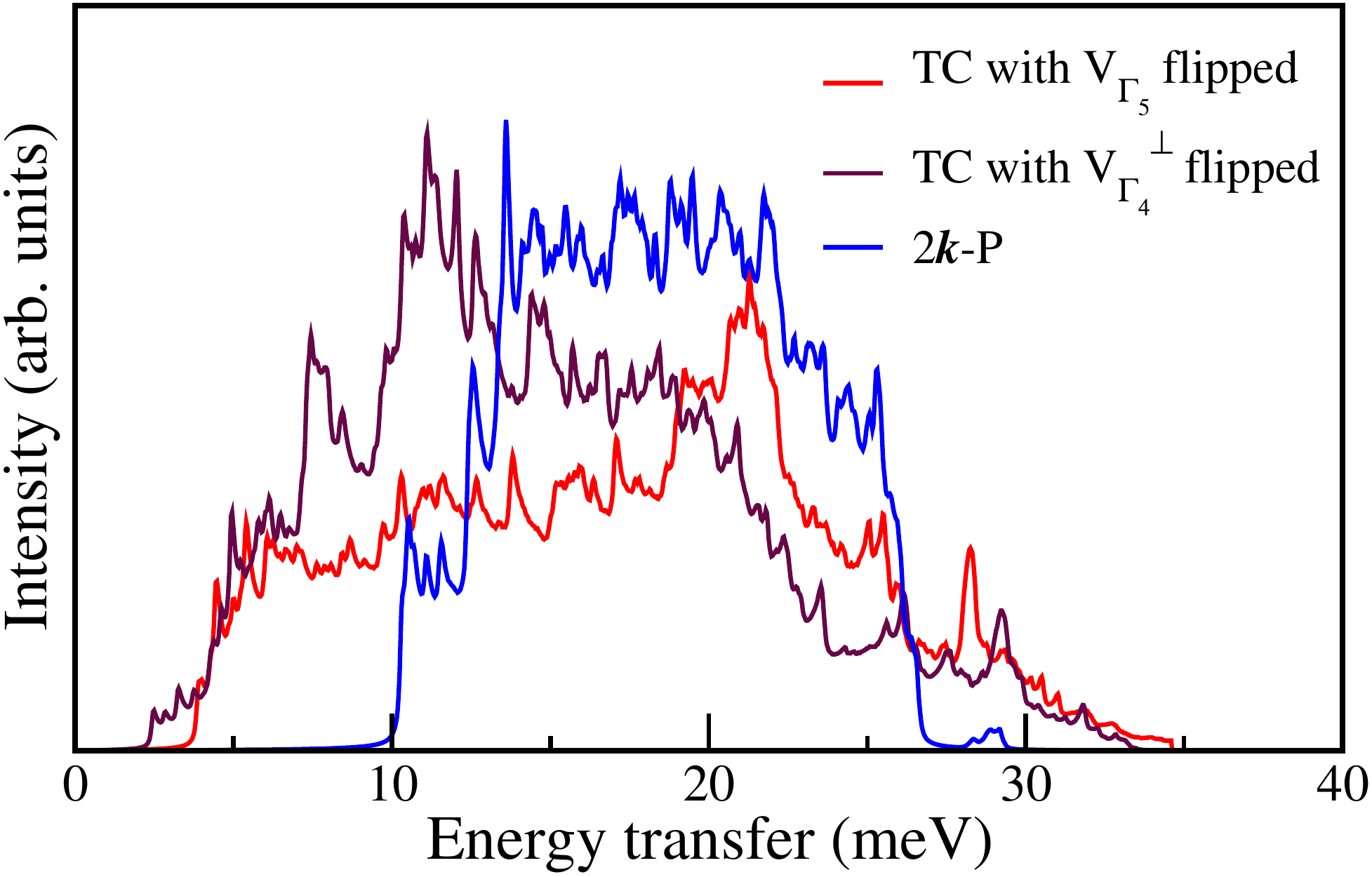} 
  		\par\end{centering}
  	\caption{Calculated \vq-integrated INS intensity for the TC structures, stabilized by flipping the sign of the corresponding DO term in the simplified Hamiltonian (eq.~2 of the main text). The INS intensity of the $2\vk$-P GS is also shown.}
  	\label{fig:INS_TC_vs_2P} 
  \end{figure}

 Basing on the analysis of the relative stability of different magnetic structures (Sec.~IIIB of the main text), any of these choices penalizes  the 2\vk-P structure; the TC one is then expected to become the GS. Indeed, by solving the resulting Hamiltonian in mean-field for BYOO, we find TC structures with collinear moments orthogonal to the [001] propagation vector (but different in-plain moment directions). The calculated INS intensity integrated over $|\vq|$ is compared in Suppl. Fig.~\ref{fig:INS_TC_vs_2P} to that for the   2\vk-P order. One sees that in the both cases, the TC spectra is still gaped, as expected for the case of non-zero DO IEI from the general arguments in the main text. However, the gap is reduced in the TC structure more than by half. Moreover, the TC spectra are significantly broader than the  2\vk-P one and agree rather badly with experimental INS intensity from Ref.~\onlinecite{Kermarrec2015}. While it  is difficult to scan over all possible combinations of the DO couplings that may give the GS TC order, these numerical experiments still give some indications of the impact of a TC GS on magnetic excitations. 
 
 \begin{figure}[!tb]
 	\begin{centering}
 		\includegraphics[width=0.57\columnwidth]{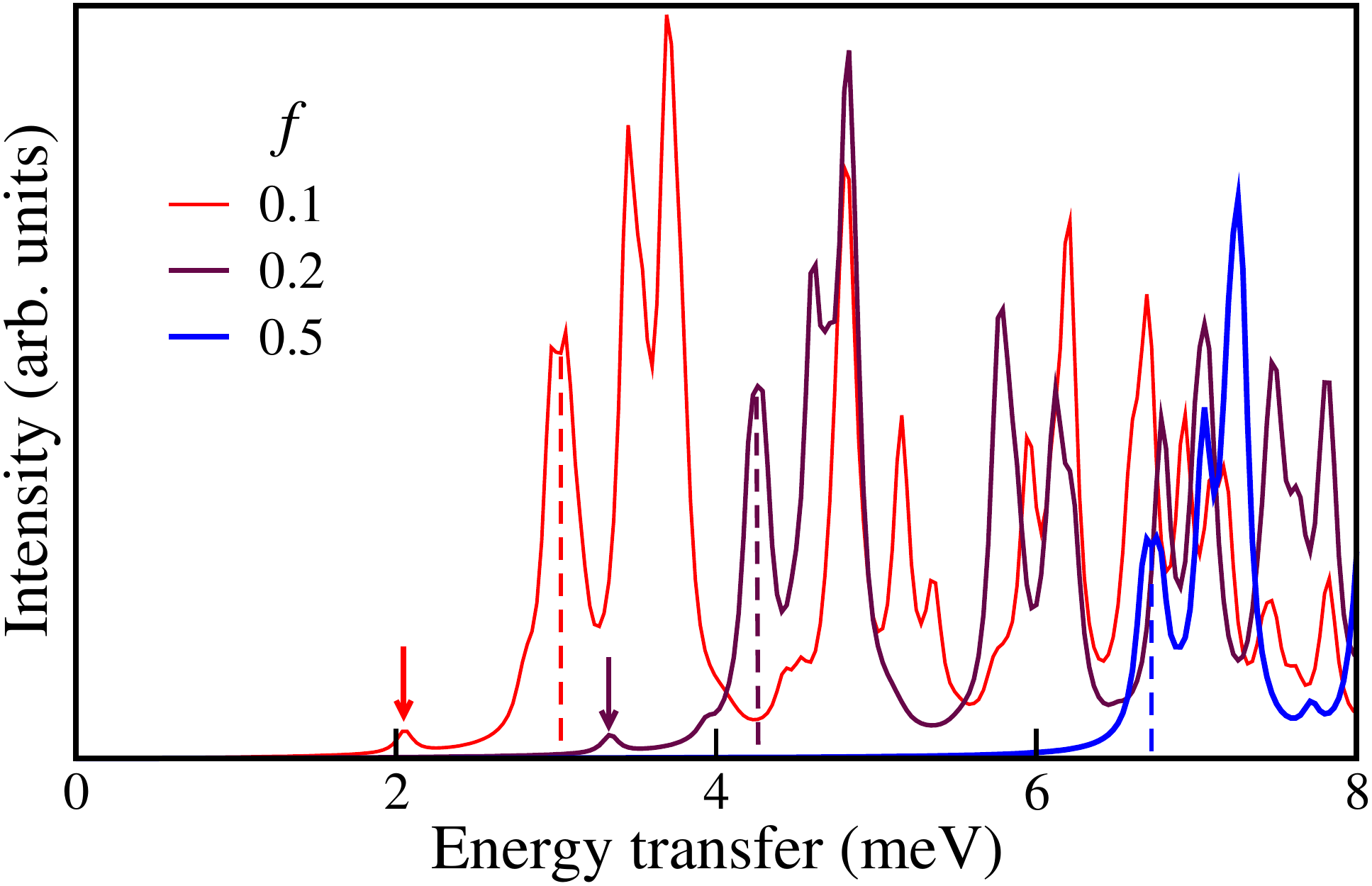} 
 		\par\end{centering}
 	\caption{Low-energy part of the calculated  BYOO INS intensity at $|\vq|=$0.75 1/\AA\ vs. the DO scaling factor $f$. The position of first major peak at the onset of spectral density  defining the excitation gap is indicated by the vertical dashed lines. The arrows indicate the weak resonance  appearing in the gap at $f < $0.5.}
 	\label{fig:INS_vs_f} 
 \end{figure}

 \section{Structure of low-energy excitations vs dipole-octupole IEI scaling}
 
 In Suppl. Fig~\ref{fig:INS_vs_f} we display the low-energy part of the INS intensity at $|\vq|=$0.75 1/\AA\  calculated at BYOO vs. the scaling factor $f$ for the DO IEI. One sees  a weak feature appearing below the onset of main spectral density at low values of $f$.

%\bibliography{bibliography}